\documentclass{aa}  
\pdfoutput=1
\usepackage[varg]{txfonts} 
\usepackage{natbib}
\usepackage{graphicx}
\usepackage{amsmath}	
\usepackage{amssymb}	
\usepackage{hyperref}
\hypersetup{
    colorlinks   = true, 
    urlcolor     = blue, 
    linkcolor    = blue, 
    citecolor   = blue 
}
\usepackage{multirow}
\usepackage{tabularx}
\usepackage[group-separator={,}]{siunitx}
\usepackage{bm}
\usepackage{xcolor}
\usepackage[export]{adjustbox}
\usepackage{url}

\usepackage[rightcaption]{sidecap}

\newcommand{\bb}[1]{\left[ #1 \right]}

\defcitealias{Li2023AA...670A.100L}{L23}
\newcommand{\Litwothree}{\citetalias{Li2023AA...670A.100L}}

\defcitealias{Conti2017MNRAS}{FC17}
\newcommand{\FC}{\citetalias{Conti2017MNRAS}}

\defcitealias{Kannawadi2019AA...624A..92K}{K19}
\newcommand{\Kone}{\citetalias{Kannawadi2019AA...624A..92K}}

\defcitealias{Asgari2021AA...645A.104A}{A21}
\newcommand{\Aone}{\citetalias{Asgari2021AA...645A.104A}}

\defcitealias{Busch2022AA...664A.170V}{vdB22}
\newcommand{\Bone}{\citetalias{Busch2022AA...664A.170V}}

\defcitealias{Dark2023arXiv230517173E}{DK23}
\newcommand{\DK}{\citetalias{Dark2023arXiv230517173E}}

\begin{document} 

   \title{KiDS-1000: Cosmology with improved cosmic shear measurements}
   \titlerunning{Improved KiDS-1000 cosmic shear}
   \authorrunning{S.-S. Li et al.}

\author{Shun-Sheng Li\inst{1}
        \and
        Henk Hoekstra\inst{1}
        \and
        Konrad Kuijken\inst{1}
        \and
        Marika Asgari\inst{2, 3}
        \and
        Maciej Bilicki\inst{4}
        \and
        Benjamin Giblin\inst{5,6}
        \and
        Catherine Heymans\inst{6,7}
        \and
        Hendrik Hildebrandt\inst{7}
        \and
        Benjamin Joachimi\inst{8}
        \and
        Lance Miller\inst{9}         
        \and
        Jan Luca van den Busch\inst{7}
        \and
        Angus H. Wright\inst{7}
        \and
        Arun Kannawadi\inst{10}
        \and
        Robert Reischke\inst{7}
        \and
        HuanYuan Shan\inst{11,12,13}
}

\institute{Leiden Observatory, Leiden University, P.O. Box 9513, 2300 RA Leiden, the Netherlands\\
            \email{ssli@strw.leidenuniv.nl}
\and
E.A Milne Centre, University of Hull, Cottingham Road, Hull, HU6 7RX, UK
\and 
Centre of Excellence for Data Science, AI, and Modelling (DAIM), University of Hull, Cottingham Road, Hull, HU6 7RX, UK
\and
Center for Theoretical Physics, Polish Academy of Sciences, al. Lotnik\'ow 32/46, 02-668 Warsaw, Poland
\and
Instituto de Ciencias del Cosmos (ICC), Universidad de Barcelona, Mart\'i i Franqu\`es, 1, 08028 Barcelona, Spain
\and
Institute for Astronomy, University of Edinburgh, Royal Observatory, Blackford Hill, Edinburgh, EH9 3HJ, UK
\and
Ruhr University Bochum, Faculty of Physics and Astronomy, Astronomical Institute (AIRUB), German Centre for Cosmological Lensing, 44780 Bochum, Germany
\and
Department of Physics and Astronomy, University College London, Gower Street, London WC1E 6BT, UK
\and
Department of Physics, University of Oxford, Denys Wilkinson Building, Keble Road, Oxford OX1 3RH, UK
\and
Department of Astrophysical Sciences, Princeton University, 4 Ivy Lane, Princeton, NJ 08544, USA
\and
Shanghai Astronomical Observatory (SHAO), Nandan Road 80, Shanghai 200030, China
\and
Key Laboratory of Radio Astronomy and Technology, Chinese Academy of Sciences, A20 Datun Road, Chaoyang District, Beijing 100101, China
\and
University of Chinese Academy of Sciences, Beijing 100049, China
}

   \date{Received 20 June 2023 / Accepted 27 September 2023}
 
  \abstract
  {
  We present refined cosmological parameter constraints derived from a cosmic shear analysis of the fourth data release of the Kilo-Degree Survey (KiDS-1000). Our main improvements include enhanced galaxy shape measurements made possible by an updated version of the \textit{lens}fit code and improved shear calibration achieved with a newly developed suite of multi-band image simulations. Additionally, we incorporated recent advancements in cosmological inference from the joint Dark Energy Survey Year 3 and KiDS-1000 cosmic shear analysis. Assuming a spatially flat standard cosmological model, we constrain $S_8\equiv\sigma_8(\Omega_{\rm m}/0.3)^{0.5} = 0.776_{-0.027-0.003}^{+0.029+0.002}$, where the second set of uncertainties accounts for the systematic uncertainties within the shear calibration. These systematic uncertainties stem from minor deviations from realism in the image simulations and the sensitivity of the shear measurement algorithm to the morphology of the galaxy sample. Despite these changes, our results align with previous KiDS studies and other weak lensing surveys, and we find a ${\sim}2.3\sigma$ level of tension with the \textit{Planck} cosmic microwave background constraints on $S_8$.}

   \maketitle
%

\section{Introduction}
\label{Sec:intro}

Weak gravitational lensing by large-scale structure, also known as cosmic shear, is a powerful technique for studying the matter distribution in the Universe without assuming a specific correlation between dark and baryonic matter~(e.g.~\citealt{Blandford1991MNRAS.251..600B,Miralda1991ApJ...380....1M,Kaiser1992ApJ...388..272K})\footnote{However, with increasing precision in weak lensing observations, the impact of baryonic processes, such as radiative cooling and feedback from star formation and active galactic nuclei, on the observed matter distribution can no longer be ignored for small-scale structures~(e.g.~\citealt{Daalen2011MNRAS.415.3649V,Semboloni2011MNRAS.417.2020S}).}. Owing to its remarkable potential in exploring the cosmic matter distribution, cosmic shear analysis has gained popularity since its first detection over 20 years ago~\citep{Bacon2000MNRAS.318..625B,Kaiser2000astro.ph..3338K,Waerbeke2000AA...358...30V,Wittman2000Natur.405..143W}. When distance information for source galaxies is also known, we can differentiate between them along the line of sight and perform a tomographic analysis, which entails reconstructing the 3D matter distribution from multiple 2D projections. This tomographic cosmic shear analysis is especially effective for constraining dark energy properties, as it sheds light on the evolution of cosmic structures~(e.g.~\citealt{Hu1999ApJ...522L..21H,Huterer2002PhRvD..65f3001H}).

Recent surveys, such as the Kilo-Degree Survey (KiDS; \citealt{Jong2013ExA....35...25D}), the Dark Energy Survey (DES; \citealt{Dark2016MNRAS.460.1270D}), and the Hyper Suprime-Cam (HSC) survey~\citep{Aihara2018PASJ...70S...4A}, primarily focus on constraining the amplitude of matter density fluctuations. Conventionally, this quantity is characterised by the parameter $S_8\equiv\sigma_8(\Omega_{\rm m}/0.3)^{0.5}$, where $\Omega_{\rm m}$ is the matter density parameter and $\sigma_8$ is the standard deviation of matter density fluctuations in spheres of radius $8h^{-1}~{\rm Mpc}$, computed using linear theory, where the Hubble constant $H_0=100h~{\rm km}~{\rm s}^{-1}~{\rm Mpc}^{-1}$. Interestingly, the $S_8$ values derived from these weak lensing surveys are consistently lower than those predicted by cosmic microwave background (CMB) observations from the \textit{Planck} satellite. 

Specifically, the latest cosmic shear analyses from KiDS~($0.759_{-0.021}^{+0.024}$; \citealt{Asgari2021AA...645A.104A}, A21 hereafter), DES~($0.759_{-0.023}^{+0.025}$; \citealt{Amon2022PhRvD.105b3514A,Secco2022PhRvD.105b3515S}), and HSC~($0.769_{-0.034}^{+0.031}$; \citealt{Li2023arXiv230400702L}; $0.776_{-0.033}^{+0.032}$; \citealt{Dalal2023arXiv230400701D}) provide $S_8$ values that are roughly $2\sigma$ lower than the \textit{Planck} predictions ($0.832\pm 0.013$; \citealt{Planck2020AA...641A...6P}) based on the standard spatially flat $\Lambda$ cold dark matter ($\Lambda$CDM) cosmological model. Most recently, a joint cosmic shear analysis of the DES Year 3 data and the fourth data release of KiDS by the two survey teams (\citealt{Dark2023arXiv230517173E}, DK23 hereafter) yields an $S_8$ constraint of $0.790^{+0.018}_{-0.014}$, which is closer to the \textit{Planck} results but still shows a level of $1.7\sigma$ difference. This mild difference in the $S_8$ constraints between the weak lensing surveys and CMB observations triggered extensive discussions from various perspectives, encompassing potential systematic errors in the data~(e.g.~\citealt{Efstathiou2018MNRAS.476..151E,Kohlinger2019MNRAS.484.3126K}), the influence of the baryonic physics~(e.g.~\citealt{Schneider2002AA...389..729S,Amon2022MNRAS.516.5355A,Preston2023arXiv230509827P}), and a potential deviation from the standard $\Lambda$CDM model~(see \citealt{Perivolaropoulos2022NewAR..9501659P} for a recent review).

Here, we focus on the control of systematics in the cosmic shear analysis, particularly those arising during the KiDS shear measurement process. Measuring lensing-induced shear from noisy pixelised galaxy images is a challenging task, complicated further by distortions caused by the point spread function (PSF) resulting from instrumental and observational conditions, as well as blending effects that arise when two or more objects are close on the sky~(see \citealt{Mandelbaum2018ARAA..56..393M} for a review). These factors can introduce significant measurement biases~(e.g. \citealt{Paulin2008AA...484...67P,Melchior2012MNRAS.424.2757M,Refregier2012MNRAS.425.1951R,Massey2013MNRAS.429..661M,Dawson2016ApJ...816...11D,Euclid2019AA...627A..59E}) and alter the selection function of the source sample, leading to selection bias~(e.g. \citealt{Hartlap2011AA...528A..51H,Chang2013MNRAS.434.2121C,Hoekstra2021AA...646A.124H}). Therefore, obtaining unbiased shear measurements requires careful calibration, which can be performed using either pixel-level image simulations~(e.g. \citealt{Miller2013MNRAS.429.2858M,Hoekstra2015MNRAS.449..685H}; \citealt{Conti2017MNRAS}, FC17 hereafter; \citealt{Samuroff2018MNRAS.475.4524S,Mandelbaum2018MNRAS.481.3170M}) or the data themselves~(e.g. \citealt{Huff2017arXiv170202600H,Sheldon2017ApJ...841...24S,Sheldon2020ApJ...902..138S}).

Additionally, in the case of large-area imaging surveys, determining the distance information for individual source galaxies depends on redshifts derived from broadband photometric observations. These photometric redshift estimates, which are subject to significant uncertainty, require careful calibration using spectroscopic reference samples~(e.g.~\citealt{Hoyle2018MNRAS.478..592H,Tanaka2018PASJ...70S...9T,Hildebrandt2021AA...647A.124H}). Furthermore, recent studies have shown that the blending of source images results in the coupling of shear and redshift biases~(e.g.~\citealt{MacCrann2022MNRAS.509.3371M}; \citealt{Li2023AA...670A.100L}, L23 hereafter). Consequently, a joint calibration of these two estimates becomes essential, which will necessitate the use of multi-band image simulations in future cosmic shear analyses.

In light of all these concerns, we implemented several improvements to the cosmic shear measurements in KiDS, as detailed in {\Litwothree}. We enhanced the accuracy of the galaxy shape measurements by using an upgraded version of the \textit{lens}fit code~\citep{Miller2007MNRAS.382..315M,Miller2013MNRAS.429.2858M,Kitching2008MNRAS.390..149K}, complemented by an empirical correction scheme that reduces PSF contamination. More notably, in {\Litwothree} we introduced SKiLLS (SURFS-based KiDS-Legacy-Like Simulations), a suite of multi-band image simulations that enables a joint calibration of shear and redshift estimates. This is an important element for the forthcoming weak lensing analysis of the complete KiDS survey, known as the KiDS-Legacy analysis (Wright et al. in prep.).

In this paper we take an intermediate step towards the forthcoming KiDS-Legacy analysis by applying the improvements from {\Litwothree} to a cosmic shear analysis based on the fourth data release of KiDS~\citep{Kuijken2019AA...625A...2K}. In contrast to previous KiDS cosmic shear analyses, which used shear calibration methods developed in {\FC} and \citet[K19 hereafter]{Kannawadi2019AA...624A..92K} based on single-band image simulations, our analysis adopted SKiLLS, marking the first instance of multi-band image simulations being used for KiDS cosmic shear analysis\footnote{{\Kone} did attempt to assign photo-$z$ estimates from data to simulations, but the actual photo-$z$ measurements were not simulated.}. We also incorporated recent advancements in cosmological inference and updated the current cosmological parameter constraints from KiDS. In particular, we updated the code for the non-linear evolution of the matter power spectrum calculation from \textsc{hmcode} to the latest \textsc{hmcode}-2020 version~\citep{Mead2021MNRAS.502.1401M}. We also investigated the impact of the intrinsic alignment (IA) model by incorporating amplitude priors inspired by \citet{Fortuna2021MNRAS.501.2983F}.

The remainder of this paper is structured as follows. In Sect.~\ref{Sec:data} we introduce and validate the updated KiDS shear catalogue, which is followed by the shear and redshift calibration in Sect.~\ref{Sec:calib}. We describe our cosmological inference method in Sect.~\ref{Sec:analysis} and present the results in Sect.~\ref{Sec:results}. Finally, we summarise the results in Sect.~\ref{Sec:sum}. 

\section{Updated weak lensing shear catalogue}
\label{Sec:data}

Our shear catalogue is based on the fourth data release of KiDS~\citep{Kuijken2019AA...625A...2K}, which combines optical observations in the $ugri$ bands from KiDS using the European Southern Observatory (ESO) VLT Survey Telescope~ \citep{Jong2013ExA....35...25D} and near-infrared observations in the $ZYJHK_s$ bands from the ESO Visible and Infrared Survey Telescope for Astronomy (VISTA) Kilo-degree INfrared Galaxy (VIKING) survey~\citep{Edge2013Msngr.154...32E}. The dataset covers $1006~{\rm deg}^2$ survey tiles and includes nine-band photometry measured using the Gaussian Aperture and PSF (\textsc{GAaP}) pipeline \citep{Kuijken2015MNRAS.454.3500K}. The photometric redshifts (photo-$z$s) for individual source galaxies were estimated using the Bayesian photometric redshift (\textsc{bpz}) code \citep{Ben2000ApJ...536..571B}. After masking, the effective area of the dataset in the Charge-Coupled Device (CCD) pixel frame is $777.4~{\rm deg}^2$ \citep{Giblin2021AA...645A.105G}. To perform the cosmic shear analysis, we divided the source sample into five tomographic bins based on the \textsc{bpz} estimates ($z_{\rm B}$). The first four bins have a spacing of $\Delta z_{\rm B}=0.2$ in the range $0.1<z_{\rm B}\leq 0.9$, while the fifth bin covers the range $0.9<z_{\rm B}\leq 1.2$, following the previous KiDS cosmic shear analyses.

\subsection{Galaxy shapes measured with the updated \textit{lens}fit}
\label{Sec:lensfit}

When preparing the shear measurements for the upcoming data release of KiDS, we upgraded the \textit{lens}fit code~\citep{Miller2007MNRAS.382..315M,Miller2013MNRAS.429.2858M,Kitching2008MNRAS.390..149K} from version $309c$ to version $321$ (see {\Litwothree} for details). The latest version includes a correction to an anisotropic error in the original likelihood sampler, which previously caused a small yet noticeable residual bias that was not related to the PSF or underlying shear~\citep{Miller2013MNRAS.429.2858M,Hildebrandt2016MNRAS.463..635H,Giblin2021AA...645A.105G}. We used the new code to re-measure the galaxy shapes, resulting in a new shear catalogue. Throughout the paper, we refer to the new shear catalogue as KiDS-1000-v2 to distinguish it from the previous KiDS-1000(-v1) shear catalogue~\citep{Giblin2021AA...645A.105G}.

The raw measurements from the \textit{lens}fit code suffer from biases primarily due to the PSF anisotropy, but also because of the object selection and weighting scheme. To address these biases, {\FC} introduced an empirical correction scheme to isotropise the original measurement weights, which was used in previous KiDS studies~(see also {\Kone}). This correction scheme mitigates the \textit{lens}fit weight biases and reduces the bias induced by the PSF anisotropy to an acceptable level. However, notable residual biases still persist~\citep{Giblin2021AA...645A.105G}. Moreover, {\Litwothree} find that the method is susceptible to variations in the sample size, posing challenges for consistent application to both data and simulations.

Therefore, a new correction scheme was introduced by {\Litwothree} that modifies both the measured ellipticities and weights to ensure the average PSF leakage, defined as the fraction of the PSF ellipticity leaking into the shear estimator, is negligible in each tomographic bin. For further details, we direct readers to {\Litwothree}. In summary, the new correction scheme first isotropises the measurement weights, then adjusts the measured ellipticities to eliminate any remaining noise bias and selection effects. We note that this correction scheme is not designed to refine the shape measurements of individual galaxies; rather, it aims to ensure that the collectively weighted shear signal is robust against PSF leakage. In this work, we applied this newly developed empirical correction to the KiDS-1000-v2 shear catalogue.

\subsection{Validation of the shear estimates}

In order to use the weak lensing shear catalogue for cosmological inference, it is crucial to first verify the accuracy of the shear estimation and ensure that the residual contamination from systematic effects is within the acceptable level for scientific analysis. To achieve this, \citet{Giblin2021AA...645A.105G} proposed a series of null-tests to assess the robustness of the KiDS-1000-v1 shear catalogue. With the updated galaxy shape measurements in the KiDS-1000-v2 catalogue, it is necessary to repeat some of these tests to confirm the reliability of the new catalogue.

As the KiDS-1000-v2 catalogue updates only the galaxy shape measurements while maintaining the established photometry and PSF models, we did not repeat tests related to photometry and PSF modelling. We started by examining the PSF leakage in the weighted \textit{lens}fit shear estimator, using the first-order systematics model proposed by \citet{Heymans2006MNRAS.368.1323H}. This model takes the form~\citep{Giblin2021AA...645A.105G}
\begin{equation}
\label{eq:OneBias}
    \epsilon_{k}^{\rm obs}=(1+m_{k})(\epsilon_{k}^{\rm int}+\gamma_{k}) + \alpha_{k}\epsilon_{k}^{\rm PSF} + c_{k}~,~~~[k=1,2]~,
\end{equation}
where $\epsilon^{\rm obs}$ denotes the measured galaxy ellipticity, $m$ is the multiplicative shear bias\footnote{Throughout this paper, we interchangeably use `multiplicative bias' and `shear bias', as our simulation-based shear calibration only addresses this parameter. Conversely, PSF leakage and the additive term are empirically corrected.}, $\epsilon^{\rm int}$ refers to the intrinsic galaxy ellipticity, $\gamma$ stands for the cosmic shear signal (which is the parameter of interest), $\alpha$ is the PSF leakage factor, and $c$ is an additive term comprising residual biases unrelated to the PSF or underlying shear. The subscript $k=1,2$ denotes the two ellipticity components. We note that we did not include PSF modelling errors in Eq.~(\ref{eq:OneBias}), as we used the same PSF model as \citet{Giblin2021AA...645A.105G}, who had already confirmed its accuracy. Assuming that $(\epsilon_{k}^{\rm int}+\gamma_{k})$ averages to zero for a large galaxy sample (a property validated with the KiDS data; see, for example, Sect.~3 in \citealt{Giblin2021AA...645A.105G}), we can determine the $\alpha$ and $c$ parameters from the data using a simple linear regression method.

\begin{figure}
  \centering
  \includegraphics[width=\hsize]{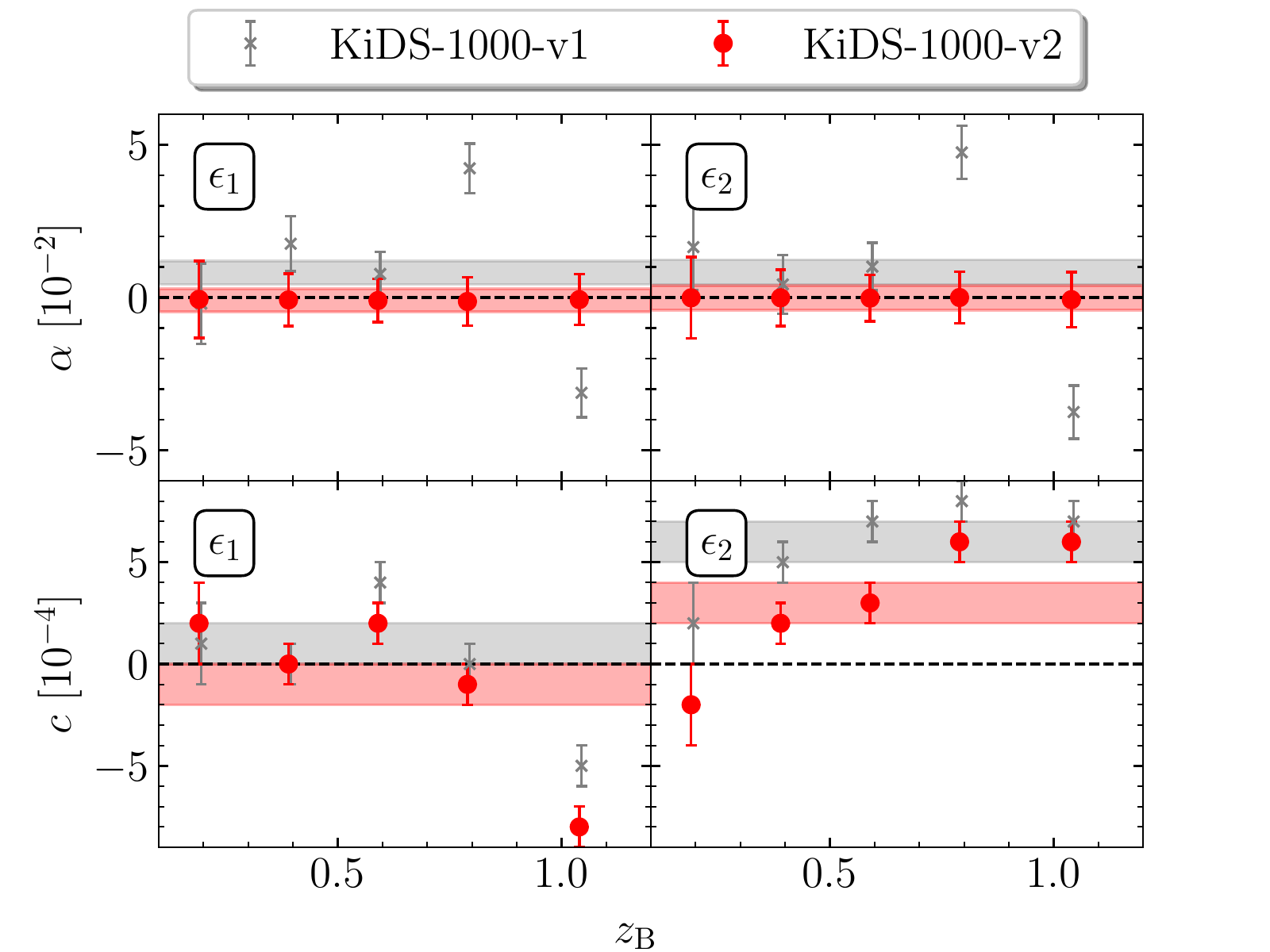}
      \caption{PSF contamination, $\alpha$ (top panels) and additive term, $c$ (bottom panels) as a function of tomographic bin labelled with the central $z_{\rm B}$ value. The measurements are obtained from a weighted linear fitting using Eq.~(\ref{eq:OneBias}). The red points represent measurements from the KiDS-1000-v2 catalogue, while the grey points show the measurements from the KiDS-1000-v1 catalogue. The red and grey bars correspond to results from the entire sample without tomographic binning.} 
         \label{fig:alpha}
  \end{figure}

Figure~\ref{fig:alpha} presents the measured PSF leakage $\alpha$ and the additive term $c$ for the KiDS-1000-v2 catalogue, alongside the measurements from the KiDS-1000-v1 catalogues for comparison. As expected, the KiDS-1000-v2 catalogue exhibits a mean $\alpha$-term consistent with zero for all redshift bins, owing to the empirical correction scheme outlined in Sect.~\ref{Sec:lensfit} (see also Sect.~4 in {\Litwothree}). The upgraded \textit{lens}fit code has reduced the overall $c_2$-term by half, reaching a level of $c_2\sim (3\pm 1)\times 10^{-4}$ for the entire sample. However, despite this improvement, the $c$ term has not been eliminated, particularly in distant tomographic bins where a small but noticeable $c$ term still persists, which was not seen in the simulations.

To correct for these residual small additive $c$-terms, we used the same empirical correction method as in previous KiDS analyses. Specifically, we subtracted the weighted average ellipticity from the observed ellipticity for each redshift bin as $\epsilon_{\rm corr}^{\rm obs}=\epsilon^{\rm obs}-\overline{\epsilon^{\rm obs}}$. Nevertheless, we caution that subtracting the mean $c$-term does not guarantee the removal of all additive biases, especially when detector-level effects, such as `charge transfer inefficiency'~(e.g.~\citealt{Rhodes2007ApJS..172..203R,Massey2010MNRAS.409L.109M}) and `pixel bounce'~(e.g.~\citealt{Toyozumi2005PASA...22..257T}), can introduce position-dependent bias patterns. Although we have detected such effects in KiDS data~\citep{Hildebrandt2020AA...633A..69H,Giblin2021AA...645A.105G}, their level does not affect the current cosmic shear analysis. More specifically, \citet{Asgari2019AA...624A.134A} show that even if current detector-level effects were increased by a factor of 10, they would not cause significant bias for KiDS-like analyses.

The cosmic shear signal is conventionally measured using the two-point shear correlation function, defined as\footnote{In this study, all measurements of the two-point shear correlation function are conducted using the \textsc{TreeCorr} code~\citep{Jarvis2004MNRAS.352..338J,Jarvis2015ascl.soft08007J}.}
\begin{equation}
\label{eq:CF}
    \hat{\xi}^{ij}_{\pm}(\theta)=\frac{\sum_{ab}w_aw_b\left[\epsilon_t^i(\bm{x}_a)\epsilon_t^j(\bm{y}_b)\pm\epsilon_{\times}^i(\bm{x}_a)\epsilon_{\times}^j(\bm{y}_b)\right]}{\sum_{ab}w_aw_b}~,
\end{equation}
where $\theta$ represents the separation angle between a pair of galaxies $(a, b)$, the tangential and cross ellipticities $\epsilon_{t, \times}$ are computed with respect to the vector $\bm{x}_a - \bm{y}_b$ that connects the galaxy pair, and the associated measurement weight is denoted by $w$. Therefore, it is crucial to examine the systematics in the two-point statistics. Following the method of \citet{Bacon2003MNRAS.344..673B}, we estimated the PSF leakage into the two-point correlation function measurement using 
\begin{equation}
\label{eq:sys}
    \xi^{\rm sys}_{\pm} = \frac{\langle\epsilon^{\rm obs}\epsilon^{\rm PSF}\rangle^2}{\langle\epsilon^{\rm PSF}\epsilon^{\rm PSF}\rangle}~,
\end{equation}
where the $\langle\cdot\rangle$ represents the correlation function. 

\begin{figure}
  \centering
  \includegraphics[width=\hsize]{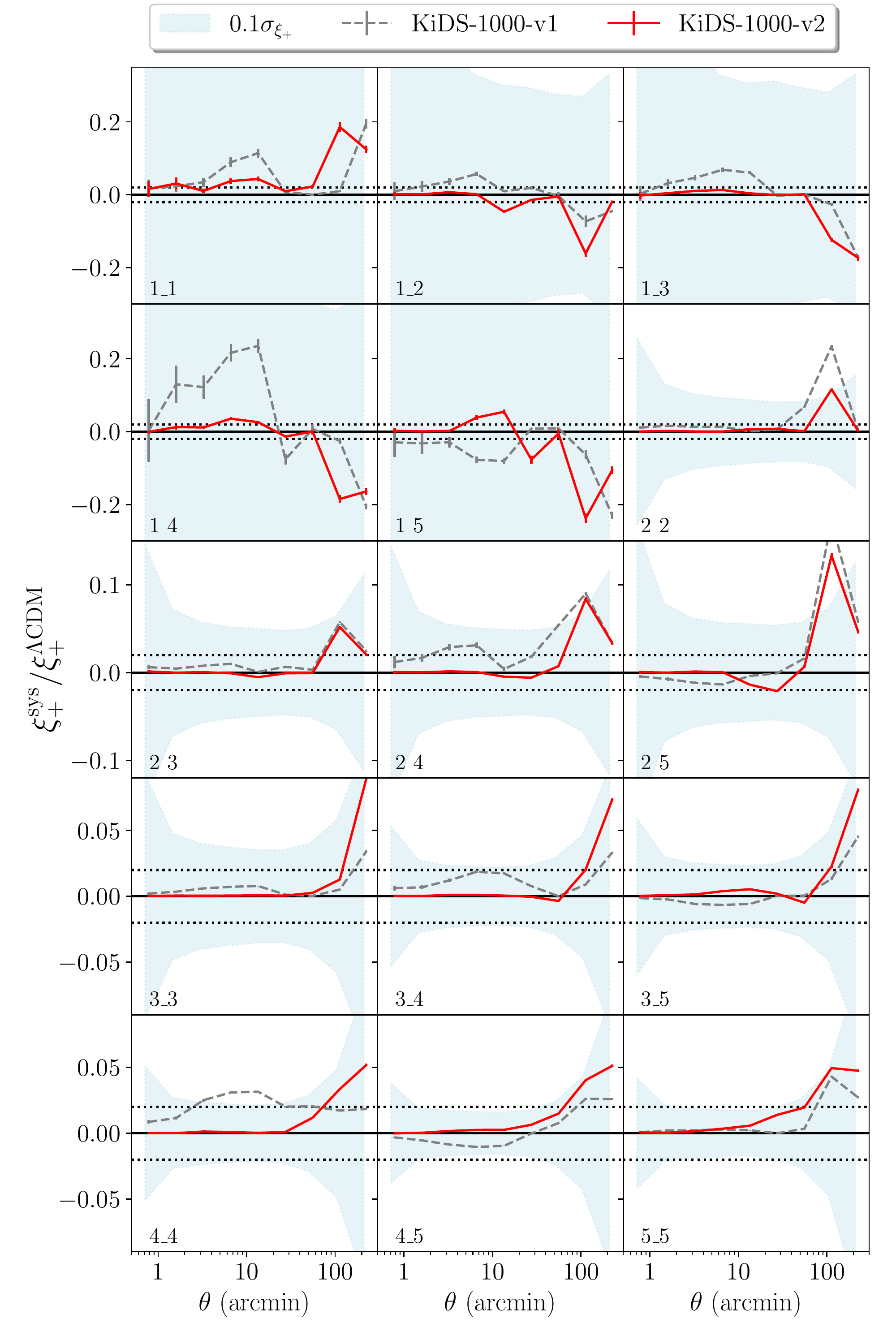}
      \caption{Ratio of the PSF contamination, $\xi_{+}^{\rm sys}$, computed using Eq.~(\ref{eq:sys}) to the predicted amplitude of the cosmic shear signal, $\xi_{+}^{\Lambda{\rm CDM}}$, across all $15$ tomographic bin combinations. The red lines depict results from the KiDS-1000-v2 catalogue, whereas the grey lines show those from the KiDS-1000-v1 catalogue. The blue shaded regions represent a range of $\pm10\%$ of the standard deviation of the measured cosmic shear signal. This deviation is determined from the covariance matrix using statistics from the KiDS-1000-v2 catalogue. The dotted horizontal lines indicate the $2\%$ level of the predicted cosmic shear signal.
      }
         \label{fig:Csys}
  \end{figure}

In Fig.~\ref{fig:Csys} we present the ratio of the measured $\xi^{\rm sys}_{+}$ to the theoretical predictions of the cosmic shear signal. The blue shaded region denotes $\pm 10\%$ of the standard deviation of the cosmic shear signal, extracted from the analytical covariance. This covariance is calculated using an independent implementation of the methodology of \citet{Joachimi2021AA...646A.129J}, and it incorporates the sample statistics of the updated catalogue. We compared the results from the KiDS-1000-v2 catalogue with those from the KiDS-1000-v1 catalogue. We observe general improvements, particularly in the high-redshift bins, where the PSF contamination is now negligible. The only exceptions are found in some large-scale bins ($\theta>60$ arcmin), where the expected fiducial cosmic shear signal is relatively small and overwhelmed by high statistical noise.

To the leading order, the weak lensing effect introduces only curl-free gradient distortions ($E$-mode signal), which makes the curl distortions ($B$-mode signal) a useful null-test for residual systematics in the shear measurement\footnote{Some higher-order effects from lensing, such as source redshift clustering~(e.g.~\citealt{Schneider2002AA...389..729S}), and IA of nearby galaxies~(e.g.~\citealt{Troxel2015PhR...558....1T,Joachimi2015SSRv..193....1J}) can also introduce $B$-mode signals. However, these contributions are expected to be negligible for current weak lensing surveys~(e.g.~\citealt{Hilbert2009AA...499...31H})}. Following the convention of KiDS~\citep{Hildebrandt2017MNRAS.465.1454H,Giblin2021AA...645A.105G}, we used the complete orthogonal sets of E/B-integrals (COSEBIs; \citealt{Schneider2010AA...520A.116S}) to measure the $B$-mode signal. The COSEBIs provide an optimal E/B separation by combining different angular scales from the $\hat{\xi}_{\pm}$ measurements. 

\begin{figure}
  \centering
  \includegraphics[width=\hsize]{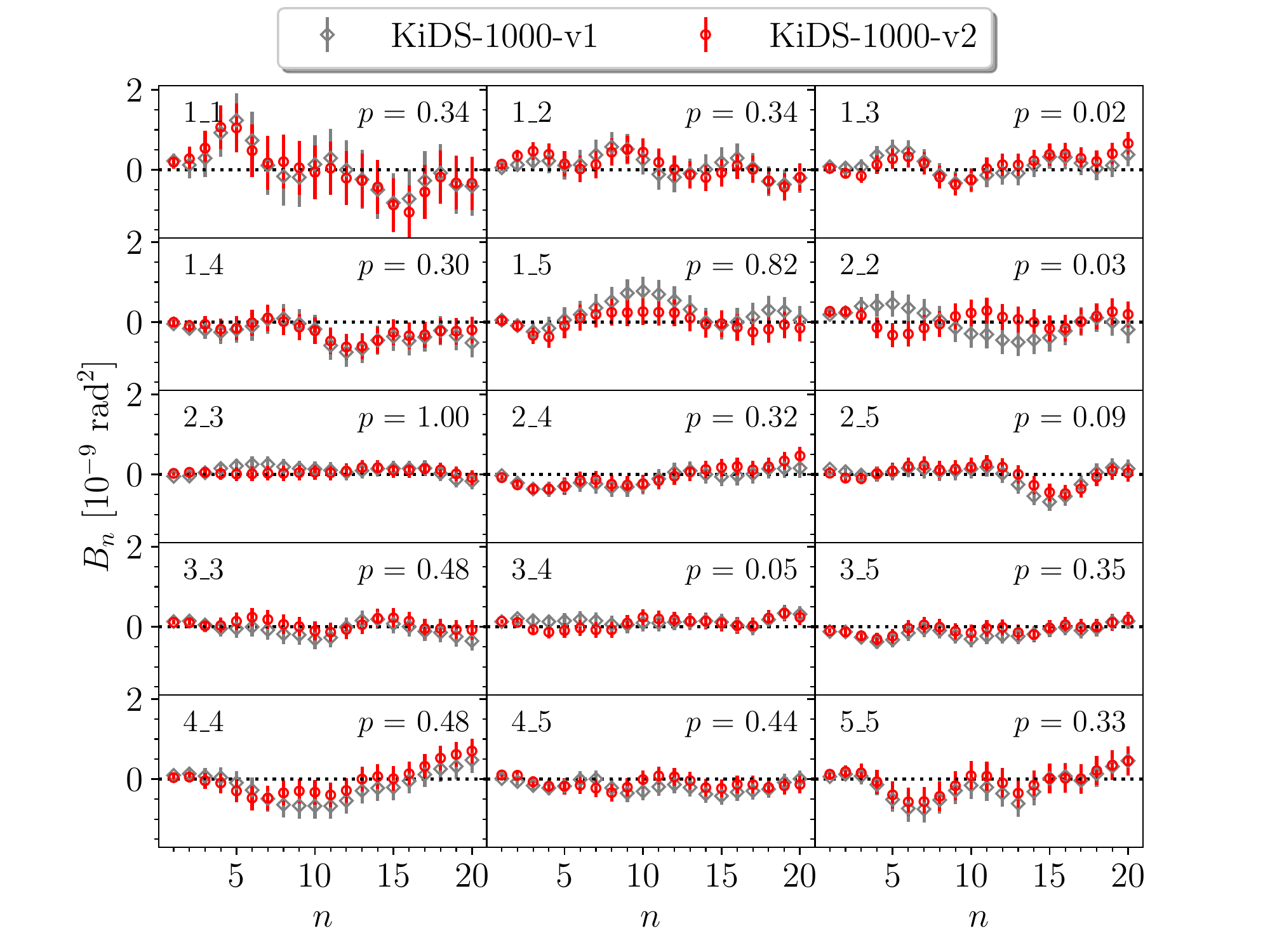}
      \caption{Measurements of the $B$-mode signals using COSEBIs for the KiDS-1000-v2 catalogue (red points) compared to the KiDS-1000-v1 catalogue (grey points). The error bars originate from the diagonal of an analytical covariance matrix, accounting solely for measurement noise. For the KiDS-1000-v2 catalogue, we re-calculated the covariance using the method introduced by \citet{Joachimi2021AA...646A.129J}, incorporating the updated statistics. The $p$-values for the KiDS-1000-v2 catalogue, shown in the top-right corner of each panel, were calculated with 20 degrees of freedom, which corresponds to the number of modes used in each correlation.}
         \label{fig:Bmode}
  \end{figure}

Figure~\ref{fig:Bmode} presents the measured $B$-mode signals for all combinations of tomographic bins in our analysis, alongside the $B$-mode measurements from the KiDS-1000-v1 catalogue for comparison. To enable a direct comparison, we used the same scale range of $(0\farcm5, 300\arcmin)$ as in \citet{Giblin2021AA...645A.105G} for calculating the COSEBIs $B$-mode\footnote{We also evaluated an alternate scale range of $(2\arcmin, 300\arcmin)$, consistent with our fiducial cosmic shear analysis. As anticipated, the $B$-mode signal was more negligible in this scenario due to reduced small-scale contamination.}. Assuming a null signal, we computed the $p$-value for each $B$-mode measurement, setting the degrees of freedom equal to the number of modes in each measurement ($n=20$). The covariance matrix, accounting only for shot noise, was estimated using an analytical model from \citet{Joachimi2021AA...646A.129J} applied to the updated catalogue. It is noteworthy that our covariance matrix differs from the one used in \citet{Giblin2021AA...645A.105G}. This is due to the changes in sample statistics resulting from the updated shape measurement code and redshift calibration relative to the KiDS-1000-v1 catalogue used in \citet{Giblin2021AA...645A.105G}. Most diagonal entries in our matrix show reduced uncertainties, ranging from a level of per cent to ten per cent. Therefore, if the absolute systematic levels are comparable between the two catalogues, our test would likely show a slight increase in the final $p$-values compared to those in \citet{Giblin2021AA...645A.105G}. As indicated in the top-right corner of each panel, the estimated $p$-values suggest that the measured $B$-mode signals align with a null signal across all bin combinations. The lowest $p$-value, $p=0.02$, was found in the cross-correlation between the first and third tomographic bins.

After conducting all these tests, we can conclude that the KiDS-1000-v2 catalogue has reduced systematics when compared to the results from the KiDS-1000-v1 catalogue. These improvements are largely attributed to the updated version of the \textit{lens}fit code, as well as the implementation of a new empirical correction scheme for PSF contamination. These results give us the confidence to use the updated catalogue for cosmological inference.

\section{Shear and redshift calibration}
\label{Sec:calib}

The main improvement in our calibration comes from the use of SKiLLS multi-band image simulations, as developed in {\Litwothree}. These simulations fuse cosmological simulations with high-quality observational data to create mock galaxies with photometric and morphological properties closely resembling real-world galaxies. The observational data used by SKiLLS, drawn from the catalogue of \citet{Griffith2012ApJS}, is identical to that used in {\Kone}. In {\Litwothree}, we developed a vine-copula-based algorithm that learns the measured morphological parameters from this catalogue and assigns them to the SURFS-Shark mock galaxies~\citep{Elahi2018MNRAS.475.5338E,Lagos2018MNRAS.481.3573L}. We verified that the learning procedure maintains the observed multi-dimensional correlations between morphological parameters, magnitude, and redshifts. Nevertheless, both the observed catalogue from \citet{Griffith2012ApJS} and the learning algorithm possess inherent limitations, resulting in unavoidable uncertainties in our simulation input catalogue. These uncertainties are addressed in our shear calibration in Sect.~\ref{Sec:biasUncertainties}.

To create KiDS+VIKING-like nine-band images, SKiLLS replicated the instrumental and observational conditions of 108 representative tiles selected from six sky pointings evenly distributed across the footprint of the KiDS fourth data release. The star catalogue was generated for each sky pointing using the \textsc{Trilegal} population synthesis code~\citep{Girardi2005AA} to account for the variation in stellar densities across the footprint. For the primary $r$-band images, on which the galaxy shapes were measured, SKiLLS included the correlated pixel noise introduced by the stacking process and the PSF variation between CCD images.

On the data processing side, SKiLLS followed the entire KiDS procedure, including object detection, PSF homogenisation, forced multi-band photometry, photo-$z$ estimation, and shape measurements. The end result is a self-consistent joint shear-redshift mock catalogue that matches KiDS observations in both shear and redshift estimates. By taking this end-to-end approach, we accounted for photo-$z$-related selection effects in our shear bias estimation and enabled redshift calibration using the same mock catalogue. While our current analysis focuses on the improvement in shear calibration, it represents an intermediate step towards the KiDS-Legacy analysis, which will implement joint shear and redshift calibrations facilitated by the SKiLLS mock catalogue.

\subsection{Calibration}
\label{Sec:calib2}

To correct for shear bias in our measurements, we adopted the method used in previous KiDS studies~({\FC}, {\Kone}). For each tomographic bin $i$, we applied an average shear bias correction factor, $\overline{m^{i}}$, which is derived by averaging the individual $m$ values of all sources within the respective tomographic bin. These individual $m$ values are determined using Eq.~(\ref{eq:OneBias}), based on simulations mapped on a grid of the \textit{lens}fit reported model signal-to-noise ratio and resolution. Here, the resolution is defined as the ratio of the PSF size to the measured galaxy size. To align our simulations more closely with the target data, we followed KiDS conventions and re-weighted the simulation estimates according to the grid of signal-to-noise ratio and resolution. Further details about the re-weighting procedure can be found in Sec.~5.1 of {\Litwothree}.

Although the averaging method addresses the noise in individual source's $m$ estimation, it does not account for correlations involving shear bias. Thus, we have $\langle[1+m^{i}(\theta')][1+m^{j}(\theta'+\theta)]\rangle= (1+\overline{m^{i}})(1+\overline{m^{j}})$, with $\theta$ and $\theta'$ representing different separation angles between galaxy pairs. To test this assumption, we directly measured $\langle[1+m^{i}(\theta')][1+m^{j}(\theta'+\theta)]\rangle$ from image simulations and compared it to $(1+\overline{m^{i}})(1+\overline{m^{j}})$. Further details on this test can be found in Appendix~\ref{Sec:CF4m}. In summary, we find a negligible difference between the two estimators, a result that falls well within the current KiDS requirements. This validates the assumption for the KiDS analysis.

Given that the updated galaxy shape measurements also lead to changes in the sample selection function, it is necessary to repeat the redshift calibration for the KiDS-1000-v2 catalogue, even though our primary focus is to improve shear calibration. To quantify the changes in galaxy samples introduced by the modifications in shape measurements from the KiDS-1000-v1 to KiDS-1000-v2 catalogues, we compared their effective number densities before applying any redshift calibration. The observed percentage differences in each tomographic bin, from low to high redshift bins, are $-1.8\%$, $-0.4\%$, $0.2\%$, $1.3\%$, and $3.2\%$. Here, negative values indicate a decrease in density from the v1 to the v2 catalogue, while positive values signify an increase. These differences are largely attributed to changes in the weighting scheme brought by the \textit{lens}fit updates, as well as the implementation of the new empirical correction scheme for PSF leakage, as discussed in Sect.~\ref{Sec:calib} and in {\Litwothree}. For this, we employed a methodology identical to the one used by \citet{Wright2020AA...640L..14W}, \citet{Hildebrandt2021AA...647A.124H} and \citet{Busch2022AA...664A.170V}, vdB22 hereafter. It is based on a direct calibration method~\citep{Lima2008MNRAS.390..118L} implemented with a self-organising map (SOM, \citealt{kohonen1982self,Masters2015ApJ...813...53M}). More information on our implementation is provided in Appendix~\ref{Sec:zCalib}, while \citet{Wright2020AA...640L..14W}, \citet{Hildebrandt2021AA...647A.124H} and {\Bone} offer more comprehensive discussions.

The SOM-based redshift calibration method uses a `gold selection' criterion to filter out sources that are not represented in the spectroscopic reference sample (see Appendix~\ref{Sec:zCalib}). However, this process influences shear biases as it alters the selection function of the final sample. To ensure a consistent estimation of shear biases, we created the SKiLLS-gold catalogue by mimicking this quality control on the SKiLLS mock catalogue, using the same SOM trained by the spectroscopic reference sample as the real data. We derived the appropriate shear bias correction factors from this SKiLLS-gold catalogue for individual tomographic bins, and present these values in Table~\ref{tab:mbias}. It is worth noting that the shear bias estimates presented in this work differ slightly from those in {\Litwothree}, which did not include the gold selection procedure. Despite this, the differences in the estimated shear biases are relatively minor across all tomographic bins, with the first tomographic bin showing the most noticeable change of 0.008.

Our fiducial results, $m_{\rm final}$, account for the impact of PSF modelling uncertainties and the `shear interplay' effect, which occurs when galaxies from different redshifts are blended together. For more details on these effects, we refer the reader to {\Litwothree} and \citet{MacCrann2022MNRAS.509.3371M}. Additionally, we provide the idealised $m_{\rm raw}$ results, which do not consider these higher-order effects. By comparing the cosmological constraints obtained from these two cases, we aim to evaluate the robustness of previous KiDS results with respect to these higher-order effects, which were not taken into account in the earlier shear calibration~({\FC}; {\Kone}).

\begin{table*}
    \centering
    \caption{Data properties for the KiDS-1000-v2 catalogue.}
    \renewcommand{\arraystretch}{1.10}
    \label{tab:mbias}
    \begin{tabular}{ccccccccc}
        \hline\hline
        Bin & Photo-$z$ range & $n_{\rm eff}~[{\rm arcmin}^{-2}]$ & $\sigma_{\epsilon,i}$ & $\delta z = z_{\rm est} - z_{\rm true}$ & $m_{\rm raw}$ & $m_{\rm final}$ & $\sigma_m$ \\
        \hline
        1   & $0.1 < z_{\rm B} \leq 0.3$ & $0.68$ & $0.27$ & $\hphantom{-}0.000 \pm 0.0096$ & $-0.023$ & $-0.021$ & $0.019$\\
        2   & $0.3 < z_{\rm B} \leq 0.5$ & $1.30$ & $0.26$ & $\hphantom{-}0.002 \pm 0.0114$ & $-0.025$ & $-0.023$ & $0.008$\\
        3   & $0.5 < z_{\rm B} \leq 0.7$ & $1.97$ & $0.28$ & $\hphantom{-}0.013 \pm 0.0116$ & $-0.013$ & $-0.015$ & $0.007$\\
        4   & $0.7 < z_{\rm B} \leq 0.9$ & $1.39$ & $0.27$ & $\hphantom{-}0.011 \pm 0.0084$ & $\hphantom{-}0.018$ & $\hphantom{-}0.015$ & $0.006$\\
        5   & $0.9 < z_{\rm B} \leq 1.2$ & $1.35$ & $0.29$ & $-0.006 \pm 0.0097$ & $\hphantom{-}0.032$ & $\hphantom{-}0.031$ & $0.006$\\
        \hline
    \end{tabular}
    \renewcommand{\arraystretch}{1.0}
    \tablefoot{Comparable summary statistics for the KiDS-1000-v1 catalogue can be found in Table 1 of {\Aone}. We note that the differences in summary statistics between our work and {\Aone} stem from both the updated \textit{lens}fit code and the enhanced redshift calibration outlined in {\Bone}. The effective number density $n_{\rm eff}$ and the ellipticity dispersion per ellipticity component $\sigma_{\epsilon, i}$ are calculated using the formulae provided in Appendix C of \citet{Joachimi2021AA...646A.129J}. The $n_{\rm eff}$ values in this table are derived from an effective area of $777.4$ square degrees in the CCD pixel frame, making them directly comparable to the values in Table 1 of {\Aone}. The correlated Gaussian redshift priors are based on the differences between the estimated and true redshifts, $\delta z = z_{\rm est} - z_{\rm true}$, as reported in {\Bone}. The priors are denoted as $\mu_i \pm \sigma_i$, where $\mu_i$ represents the mean shift and $\sigma_i$ corresponds to the square root of the covariance matrix's diagonal elements. The $m_{\rm raw}$ results are derived from idealised constant shear simulations, while the $m_{\rm final}$ results, our fiducial outcomes, include corrections for the shear-interplay effect and PSF modelling bias. Statistical uncertainties, determined by the simulation volume, are directly computed from the fiducial simulations and denoted as $\sigma_m$.}
\end{table*}

\subsection{Calibration uncertainties}
\label{Sec:biasUncertainties}

Systematic uncertainties arising from redshift and shear calibrations can propagate into cosmological analyses, potentially leading to biased results. Therefore, it is crucial to adequately address these uncertainties in the analysis. In this section we outline our approach to managing these calibration uncertainties.

The uncertainties in redshift calibration were addressed by introducing an offset parameter for the estimated mean redshift of galaxies in each tomographic bin. This offset parameter, described as correlated Gaussian priors, serves as a first-order correction to both the statistical and systematic uncertainties associated with redshift calibration. Table~\ref{tab:mbias} lists the exact values for these parameters, which we obtained from {\Bone} and \citet{Hildebrandt2021AA...647A.124H}. They determined these prior values using spectroscopic and KiDS-like mock data generated by \citet{Busch2020AA...642A.200V}. We consider the current priors to be conservative enough to account for any potential changes in the redshift biases from KiDS-1000-v1 to KiDS-1000-v2, given that both catalogues use the same photometric estimates. However, for the forthcoming KiDS-Legacy analysis, we plan to re-estimate these values based on the new SKiLLS mock data.

We improved our approach to handling uncertainties related to the shear calibration. In {\Litwothree}, nominal uncertainties were proposed for each tomographic bin based on sensitivity analyses. This aimed to ensure the robustness of the shear calibration within the specified uncertainties, but at the cost of reducing statistical power. In this work, we aim to improve this approach by separately accounting for the statistical and systematic uncertainties within the shear calibration.

The statistical uncertainties, as presented in Table~\ref{tab:mbias}, are computed directly from simulations and are limited only by the volume of the simulations, which can be increased with more computing resources\footnote{However, the finite volume of the input galaxy sample prevents an indefinite increase.}. These uncertainties are also easily propagated into the covariance matrix for cosmological inference. Although increasing the simulation volume could, in principle, reduce these uncertainties, we find that the current values already comfortably meet the KiDS requirements; thus, further efforts in this direction were considered unnecessary. 

If the SKiLLS simulations perfectly match KiDS data, these statistical uncertainties would be the only contribution to the final uncertainty from the shear calibration. However, since our simulations are not a perfect replica of the real observations, residual shear biases may still be present in the data even after calibration. These biases, referred to as systematic uncertainties, are typically the primary source of error in shear calibration. Increasing the simulation volume cannot improve these uncertainties as they are determined by the realism of the image simulations. The level of these uncertainties can only be roughly estimated through sensitivity analyses.

Since the systematic residual shear biases directly scale the data vector, accurately quantifying their impact using the covariance matrix is challenging. Therefore, we used a forward modelling approach to capture the impact of these systematic uncertainties. Instead of incorporating these uncertainties into the covariance matrix, we examined how the final estimates of the cosmological parameters change due to the shift in signals caused by the systematic residual shear biases. This forward modelling approach can be easily implemented using simple optimisation algorithms since the shift is small, and the covariance remains unchanged. More details on how to determine residual shear biases and implement the forward modelling approach are provided in Appendix~\ref{Sec:Dm}.

\section{Cosmological inference}
\label{Sec:analysis}

The cosmological inference in this study largely aligns with the approach used in the KiDS-1000-v1 analyses~({\Aone}; {\Bone}), with minor modifications primarily influenced by the recent joint DES Y3+KiDS-1000 cosmic shear analysis~(\DK). In this section, we outline the configurations and reasoning behind these choices in our fiducial analysis. For certain notable changes, we also conducted extended analysis runs with different configurations to evaluate the impact of these modifications. Our analysis code is publicly accessible\footnote{\url{https://github.com/lshuns/CSK1000LF321}}.

Our code builds upon the \texttt{Cat\_to\_Obs\_K1000\_P1}\footnote{\url{https://github.com/KiDS-WL/Cat_to_Obs_K1000_P1}} and the KiDS Cosmology Analysis Pipeline (KCAP)\footnote{\url{https://github.com/KiDS-WL/kcap}} infrastructure, as developed in \citet{Giblin2021AA...645A.105G}, \citet{Joachimi2021AA...646A.129J}, {\Aone}, \citet{Heymans2021AA...646A.140H}, and \citet{Troster2021AA...649A..88T}. The \texttt{Cat\_to\_Obs\_K1000\_P1} pipeline converts KiDS shear and redshift measurements into various second-order statistics, with the assistance of the \textsc{TreeCorr} code \citep{Jarvis2004MNRAS.352..338J,Jarvis2015ascl.soft08007J}. Meanwhile, KCAP estimates cosmological parameters using the \textsc{CosmoSIS} framework, which bridges the likelihood function calculation pipelines with sampling codes~\citep{Zuntz2015AC....12...45Z}.

We measured the shear field using COSEBIs~\citep{Schneider2010AA...520A.116S}). As reported by \citet{Asgari2020AA...634A.127A}, COSEBIs offer enhanced robustness against small-scale effects on the shear power spectrum, which primarily stem from complex baryon feedback. Furthermore, we accounted for baryon feedback when modelling the matter-matter power spectrum using \textsc{hmcode}-2020~\citep{Mead2021MNRAS.502.1401M} within the \textsc{camb} framework with the version 1.4.0~\citep{Lewis2000ApJ...538..473L,Howlett2012JCAP...04..027H}.

\textsc{hmcode}-2020, an updated version of \textsc{hmcode}~\citep{Mead2015MNRAS.454.1958M,Mead2016MNRAS.459.1468M}, models the non-linear matter-matter power spectrum, incorporating the influence of baryon feedback through an enhanced halo-model formalism. This updated model is empirically calibrated using hydrodynamical simulations, following a more physically informed approach. Unlike its predecessor calibrated with OWLS hydrodynamical simulations~\citep{Daalen2011MNRAS.415.3649V}, this newer version uses the updated BAHAMAS hydrodynamical simulations for calibration~\citep{McCarthy2017MNRAS.465.2936M}. These simulations, in turn, are calibrated to reproduce the observed galaxy stellar mass function and the hot gas mass fractions of groups and clusters. This calibration ensures that the simulation accurately reflects the impact of feedback on the overall distribution of matter (refer to \citealt{McCarthy2017MNRAS.465.2936M} for further details). Furthermore, \textsc{hmcode}-2020 improves the modelling of baryon-acoustic oscillation damping and massive neutrino treatment, achieving an improved accuracy of $2.5\%$ (compared to the previous version's $5\%$) for scales $k<10h~{\rm Mpc}^{-1}$ and redshifts $z<2$~\citep{Mead2021MNRAS.502.1401M}.

The model incorporates a single-parameter variant, $T_{\rm AGN}$, representing the heating temperature of active galactic nuclei (AGNs). Higher $T_{\rm AGN}$ values correspond to more intense AGN feedback, leading to a lower observed matter power spectrum. Following {\DK}, we used a uniform prior on $\log_{10}(T_{\rm AGN})$ that ranged from $7.3$ to $8.0$. This choice was motivated by the findings from the BAHAMAS hydrodynamical simulations~\citep{McCarthy2017MNRAS.465.2936M,Daalen2020MNRAS.491.2424V}.

Given the characteristics of COSEBIs and the implementation of the \textsc{hmcode}, the KiDS-1000-v1 analyses included small-scale measurements down to $\theta_{\rm min}=0\farcm 5$. This strategy was, however, re-evaluated in {\DK}, who suggest more stringent scale cuts for the KiDS COSEBIs data vector, determined by the baryon feedback mitigation strategy proposed by \citet{Krause2021arXiv210513548K}. 
Following this recommendation, we applied a scale cut of $\theta_{\rm min}=2\arcmin$ in our fiducial analysis.

We used the non-linear linear alignment (NLA) model to describe the IA of galaxies. This model combines the linear alignment model with a non-linear power spectrum and contains a single free parameter $A_{\rm IA}$ to describe the amplitude of IA signals~\citep{Hirata2004PhRvD..70f3526H,Bridle2007NJPh....9..444B}. It is also common to include a power law, with an index denoted as $\eta_{\rm IA}$, to capture potential redshift evolution of the IA strength. To distinguish it from the redshift-independent NLA model, we refer to this variant as the NLA-$z$ model.

In line with previous KiDS analyses, we took the redshift-independent NLA model as our fiducial choice since introducing $\eta_{\rm IA}$ has a minimal effect on the primary $S_8$ constraint (\Aone) and since current direct observations of IA signals show little evidence of substantial redshift evolution (e.g. \citealt{Joachimi2011AA...527A..26J,Singh2015MNRAS.450.2195S,Johnston2019AA...624A..30J,Fortuna2021AA...654A..76F,Samuroff2022arXiv221211319S}). However, \citet{Fortuna2021MNRAS.501.2983F} suggest that the selection of galaxy samples resulting from the redshift binning may introduce a detectable redshift variation in the IA signal, although its impact remains negligible for current weak lensing analyses. To assess the impact of $\eta_{\rm IA}$ on our results, we performed an extended run using the NLA-$z$ model, following the same prior selection as in {\DK}.

The KiDS-1000-v1 analyses adopted a broad and uninformative prior for $A_{\rm IA}$, ranging from $[-6, 6]$, considering that the data can constrain it and that an incorrect informative prior could bias the final cosmological results. Although uncertainties regarding IA signals remain large, recent developments in the field have improved our knowledge of the expected IA signal strength. For instance, \citet{Fortuna2021MNRAS.501.2983F} used a halo model formalism, incorporating results from the latest direct IA measurements, and predicted $A_{\rm IA}=0.44\pm 0.13$ for the redshift-independent NLA model targeted for KiDS-like mixed-colour lensing samples\footnote{\citet{Fortuna2021MNRAS.501.2983F} also examined the NLA-$z$ model under similar conditions, but found the fits were predominantly driven by the low-redshift bins, resulting in less accurate recovery of large-scale alignments at high redshifts.}. This prediction aligns well with the constraints from recent cosmic shear analyses~({\Aone}; \citealt{Secco2022PhRvD.105b3515S,Li2023arXiv230400702L,Dalal2023arXiv230400701D}). Moreover, recent studies revealed that other nuisance parameters in such analyses, especially those related to redshift calibration uncertainties, can result in misleading $A_{\rm IA}$ values~\citep{Hikage2019PASJ...71...43H,Wright2020AA...640L..14W,Li2021AA...646A.175L,Fischbacher2023JCAP...01..033F}. 

Given these considerations, we consider it necessary to explore the prior for the $A_{\rm IA}$ parameter. As an initial step towards a fully informed $A_{\rm IA}$ approach, we began by simply narrowing the previously broad prior, leaving a more comprehensive exploration of the IA model setups for the forthcoming KiDS-Legacy analysis. In our fiducial analysis, we chose a flat yet narrower prior of $[-0.2, 1.1]$, which corresponds to the $5\sigma$ credible region of predictions by \citet{Fortuna2021MNRAS.501.2983F}. We note that our new prior will not significantly impact the sampling results, provided that the final posterior distributions fall within the set prior range. For comparison purposes, we also conducted a test run using the wider $[-6, 6]$ prior.

Sampling the high-dimensional posterior distribution is a challenging task. In the KiDS-1000-v1 analyses, an ellipsoidal nested sampling algorithm, \textsc{MultiNest}~\citep{Feroz2009MNRAS.398.1601F}, was used. However, recent studies demonstrated that \textsc{MultiNest} systematically underestimates the $68\%$ credible intervals for $S_8$ by about $10\%$ in current weak lensing analyses~(\citealt{Lemos2023MNRAS.521.1184L}; {\DK}; \citealt{Li2023arXiv230400702L}). A promising alternative is the sliced nested sampling algorithm, \textsc{PolyChord}~\citep{Handley2015MNRAS.450L..61H,Handley2015MNRAS.453.4384H}. It provides more accurate estimates of parameter uncertainties, making it our choice for the main analysis. However, it is worth noting that \textsc{PolyChord} is nearly five times slower than \textsc{MultiNest}. Consequently, we retained \textsc{MultiNest} for testing purposes. For our sampler settings, we followed {\DK}, adopting parameters $n_{\rm live}$=500, $n_{\rm repeats}$=60 and tolerance=0.01 for \textsc{PolyChord}; and $n_{\rm live}$=1000, efficiency=0.3, tolerance=0.01, and constant efficiency=False for \textsc{MultiNest}.

When presenting point estimates and associated uncertainties for parameter constraints, we adhere to the recommendations of \citet{Joachimi2021AA...646A.129J}. We derived our best-fit point estimates from the parameter values at the maximum a posterior (MAP). Given that the MAP reported by the sampling code can be affected by noise due to the finite number of samples, we enhanced the precision of the MAP by conducting an additional local optimisation step. This process initiates from the MAP reported by the sampling code and utilises the Nelder-Mead minimisation method~\citep{Nelder10.1093/comjnl/7.4.308}, a method also employed by {\Aone}. To represent uncertainties linked to these estimates, we computed the $68\%$ credible interval based on the projected joint highest posterior density (PJ-HPD) region. This hybrid approach is more robust against projection effects stemming from high-dimensional asymmetric posterior distributions than traditional 1D marginal summary statistics (refer to Sect.~6 in \citealt{Joachimi2021AA...646A.129J} for a comprehensive discussion). To facilitate comparison with results from other surveys, we also provide constraints based on the traditional mean and maximum of the 1D marginal posterior, along with their respective $68\%$ credible intervals.

It is worth noting that, as systematic uncertainties from shear calibration are excluded in the construction of our covariance matrix (see Sect.~\ref{Sec:biasUncertainties}), the uncertainties derived from the main sampling chains do not fully account for the true uncertainties. To compensate for the additional uncertainties arising from residual shear biases, we employed a forward modelling approach. This method involves shifting the data vector and subsequently the likelihood, based on the estimated residual shear biases, followed by recalculating the MAP. As the adjustment is minor and the covariance matrix remains static, it is not necessary to re-sample the posterior distribution. Instead, we simply needed to repeat the previously mentioned local optimisation step. Starting with the original MAP and using the updated likelihood, we can determine the new MAP corresponding to each shift in the data vector. The variation in these MAP estimates represents additional uncertainties introduced by the systematic uncertainties arising from shear calibration. Further details on this process can be found in Appendix~\ref{Sec:Dm}.

Table~\ref{tab:priors} summarises the model parameters and their priors as used in our fiducial analysis. These parameters can be broadly classified into two categories: the first category includes five cosmological parameters, which describe the spatially flat $\Lambda$CDM model we employed. We fixed the sum of the neutrino masses to a value of $0.06~{\rm eV}~c^{-2}$, where $c$ is the speed of light. This choice is based on the \citet{Hildebrandt2020AA...633A..69H} finding of the negligible influence of neutrinos on cosmic shear analyses. The second category encompasses three nuisance parameters, accounting for astrophysical and measurement uncertainties as previously discussed. We note that all parameters, with the exception of $T_{\rm AGN}$ and $A_{\rm IA}$, retain the same priors as those used in the KiDS-1000-v1 cosmic shear analyses. The $T_{\rm AGN}$ parameter replaces the previous baryon feedback amplitude parameter associated with the preceding version of \textsc{hmcode}, while the $A_{\rm IA}$ parameter adopts a narrower prior for reasons previously discussed.

\begin{table}
\caption{Fiducial model parameters and their priors.}              
\label{tab:priors}      
\centering                                      
\begin{tabular}{lll}          
\hline\hline                        
Parameter & Symbol & Prior \\    
\hline                                   
Density fluctuation amp. & $S_8$ & $\bb{0.1,\,1.3}$ \\
CDM density & $\omega_{\rm c}$ & $\bb{0.051,\,0.255}$ \\
Baryon density & $\omega_{\rm b}$ & $\bb{0.019,\,0.026}$ \\
Hubble constant & $h$ & $\bb{0.64,\,0.82}$ \\
Scalar spectral index & $n_{\rm s}$ & $\bb{0.84,\,1.1}$ \\
\hline
AGN heating temperature & $\log_{10}(T_{\rm AGN}[{\rm K}])$ & $\bb{7.3,\,8.0}$\\
Intrinsic alignment amp. & $A_{\rm IA}$ & $\bb{-0.2,\,1.1}$ \\
Redshift offsets & $\delta_z$ & ${\cal N}(\vec{\mu};\vec{\sigma^2})$ \\
\hline
\end{tabular}
\tablefoot{The first section lists the primary cosmological parameters describing the $\Lambda$CDM model assumed, while the second section contains nuisance parameters related to baryon feedback, intrinsic alignments, and redshift biases. The values in square brackets indicate the limits of top-hat priors. The notation ${\cal N}(\vec{\mu};\vec{\sigma^2})$ refers to a normal prior with mean $\vec{\mu}$ and (co-)variance $\vec{\sigma^2}$, as specified in Table~\ref{tab:mbias}.}
\end{table}

\section{Results}
\label{Sec:results}

In this section we present our cosmological parameter constraints and evaluate the robustness of our findings against a variety of systematic uncertainties. We begin by presenting the outcomes from our fiducial analysis in Sect.~\ref{Sec:resFid}. We then assess the impact of shear biases in Sect.~\ref{Sec:resShear}, by quantifying the shifts in final constraints resulting from different shear bias scenarios. This highlights the main development of our work. Additionally, since we implemented several changes to the cosmological inference pipeline, we evaluate the effects of these adjustments by comparing results from multiple setup variations in Sect.~\ref{Sec:resPrevious}.

\subsection{Fiducial analysis results}
\label{Sec:resFid}

Our fiducial model has a total of twelve free parameters: five are cosmological parameters specifying the spatially flat $\Lambda$CDM model with a fixed total neutrino mass, and the remaining seven are nuisance parameters addressing astrophysical and redshift calibration uncertainties, as detailed in Sect.~\ref{Sec:analysis}. However, not all of these parameter are constrained by the cosmic shear analysis. In this section, we focus on the primary parameters that our analysis constrains. Meanwhile, the posterior distributions for all free parameters are displayed as contour plots in Appendix~\ref{Sec:allPara} for reference.

Table~\ref{tab:res} provides the point estimates along with their corresponding $68\%$ credible intervals for the primary parameter as constrained by our fiducial analysis using the \textsc{PolyChord} sampling code. We display results using three summary statistics: MAP and PJ-HPD, the mean of the 1D marginal posterior, and the maximum of the 1D marginal. As discussed in {\DK}, each of these approaches has its own advantages and limitations. Specifically, the accurate determination of MAP and PJ-HPD can be challenging, while marginal constraints for multi-dimensional posteriors are prone to projection effects. Aligning with the KiDS convention, we chose the MAP and PJ-HPD constraints as our headline results, but caution against direct comparisons with results from other surveys that might use different summary statistics. The uncertainties we report include additional contributions from the systematic uncertainties associated with our shear calibration, as detailed in Sect.~\ref{Sec:resShear}. These additional uncertainties are overall small compared to the main sampling uncertainties, so when plotting the posterior distributions or conducting extended runs for test purposes, we did not incorporate these uncertainties.

\begin{table}
    \centering
    \caption{Primary parameter constraints from our fiducial analysis, based on the KiDS-1000-v2 catalogue, as determined using the \textsc{PolyChord} sampling code.}
    \label{tab:res}
    \renewcommand{\arraystretch}{1.5}
    \begin{tabular}{lccc}
        \hline\hline
        Parameter & MAP \& PJ-HPD & \multicolumn{2}{c}{Marginal}\\
        & & Mean & Maximum\\
        \hline
         $S_8$ & $0.776_{-0.027-0.003}^{+0.029+0.002}$ & $0.765_{-0.023}^{+0.029}$ & $0.769_{-0.029}^{+0.027}$ \\
         $\Omega_{\rm m}$ & $0.259_{-0.064-0.001}^{+0.115+0.001}$ & $0.302_{-0.115}^{+0.062}$ & $0.273_{-0.088}^{+0.102}$ \\
         $\sigma_{8}$ & $0.835_{-0.158-0.003}^{+0.151+0.002}$ & $0.791_{-0.163}^{+0.124}$ &$0.752_{-0.129}^{+0.173}$ \\
         \hline
         $A_{\rm IA}$ & $0.348_{-0.322-0.011}^{+0.350+0.009}$ & $0.400_{-0.339}^{+0.330}$ & $0.397_{-0.346}^{+0.340}$\\
        \hline
    \end{tabular}
    \renewcommand{\arraystretch}{1}
    \tablefoot{Our headline results, based on the MAP and PJ-HPD statistics, include additional uncertainties that account for systematic uncertainties within the shear calibration. These uncertainties, originating from minor deviations from realism in the image simulations and the shear measurement algorithm's sensitivity to the morphology of the galaxy sample, are estimated using a forward modelling approach (as detailed in Sect.~\ref{Sec:resShear}). On the other hand, the statistical uncertainties within the shear calibration, determined by the simulation volume, are folded into the main uncertainties through their inclusion in the covariance matrix used for the cosmological inference. The mean-marginal is determined through \texttt{postprocess} within \textsc{CosmoSIS} using the default settings~\citep{Zuntz2015AC....12...45Z}; while the max-marginal is calculated using the \texttt{ChainConsumer} with the settings of statistics=`max' and kde=1.0~\citep{Hinton2016JOSS....1...45H}. The indicated uncertainties correspond to the $68\%$ credible intervals.}
\end{table}

\begin{figure}
  \centering
  \includegraphics[width=\hsize]{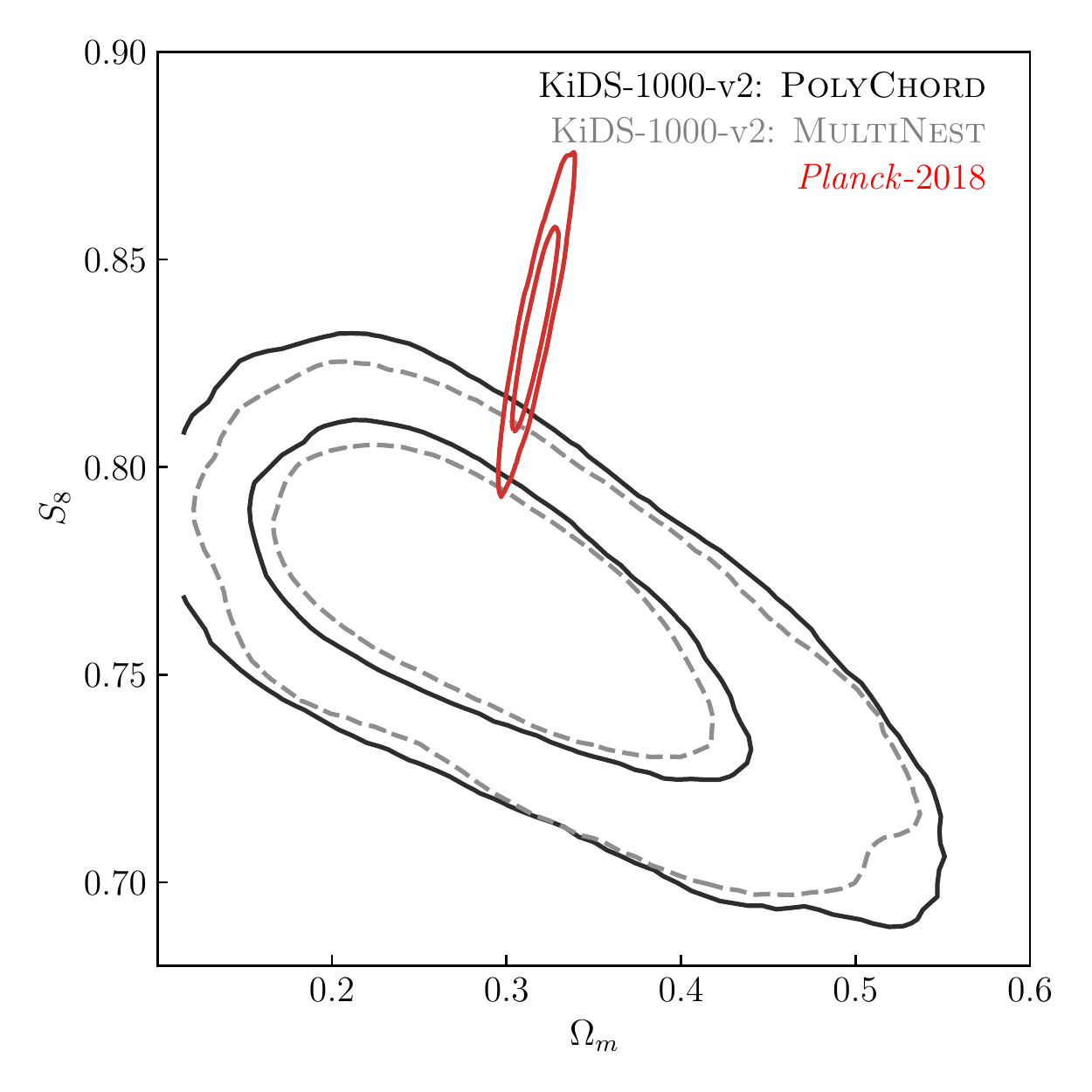}
      \caption{Comparison of projected 2D posterior distributions for the parameters $\Omega_{\rm m}$ and $S_8$ as derived from our fiducial setups using two sampling codes --- \textsc{PolyChord} (solid black line) and \textsc{MultiNest} (dashed grey line) --- against the \textit{Planck}-2018 results (solid red line). The contours correspond to the $68\%$ and $95\%$ credible intervals and are smoothed using a Gaussian kernel density estimation (KDE) with a bandwidth scaled by a factor of $1.5$, made possible by the \texttt{ChainConsumer} package~\citep{Hinton2016JOSS....1...45H}.} 
         \label{fig:S8OmFiducial}
  \end{figure}

Figure~\ref{fig:S8OmFiducial} shows the projected 2D posterior distributions for the parameters $\Omega_{\rm m}$ and $S_8$, as derived from our fiducial setups employing \textsc{PolyChord} and \textsc{MultiNest}. We see that \textsc{MultiNest} results yield a roughly $10\%$ narrower width of the posterior distribution compared to \textsc{PolyChord}, aligning with previous findings~(\citealt{Lemos2023MNRAS.521.1184L}; {\DK}; \citealt{Li2023arXiv230400702L}). However, as expected, the results from the two sampling codes show consistency in terms of best-fit values. 

In addition, we compared our cosmic shear results with those from the CMB analysis by the \textit{Planck} satellite, using their baseline $\Lambda$CDM chains with the \texttt{Plik} likelihood from their most recent \textit{Planck}-2018 results~\citep{Planck2020AA...641A...6P}. More specifically, we used their constraints based on the auto power spectra of temperature (TT), of $E$-modes (EE), and their cross-power spectra (TE), excluding CMB lensing signals. An offset is evident between our results and those from \textit{Planck}-2018. Adopting the Hellinger distance tension metric~(\citealt{Beran10.1214/aos/1176343842}; \citealt{Heymans2021AA...646A.140H}; {\DK}), we detect a $2.35\sigma$ tension in the constrained $S_8$ values. For the constrained parameter set ($S_8$, $\Omega_{\rm m}$), a similar level of tension, $2.30\sigma$, was found using the Monte Carlo exact parameter shift method~(\citealt{Raveri2020PhRvD.101j3527R}; {\DK}).

\begin{figure}
  \centering
  \includegraphics[width=\hsize]{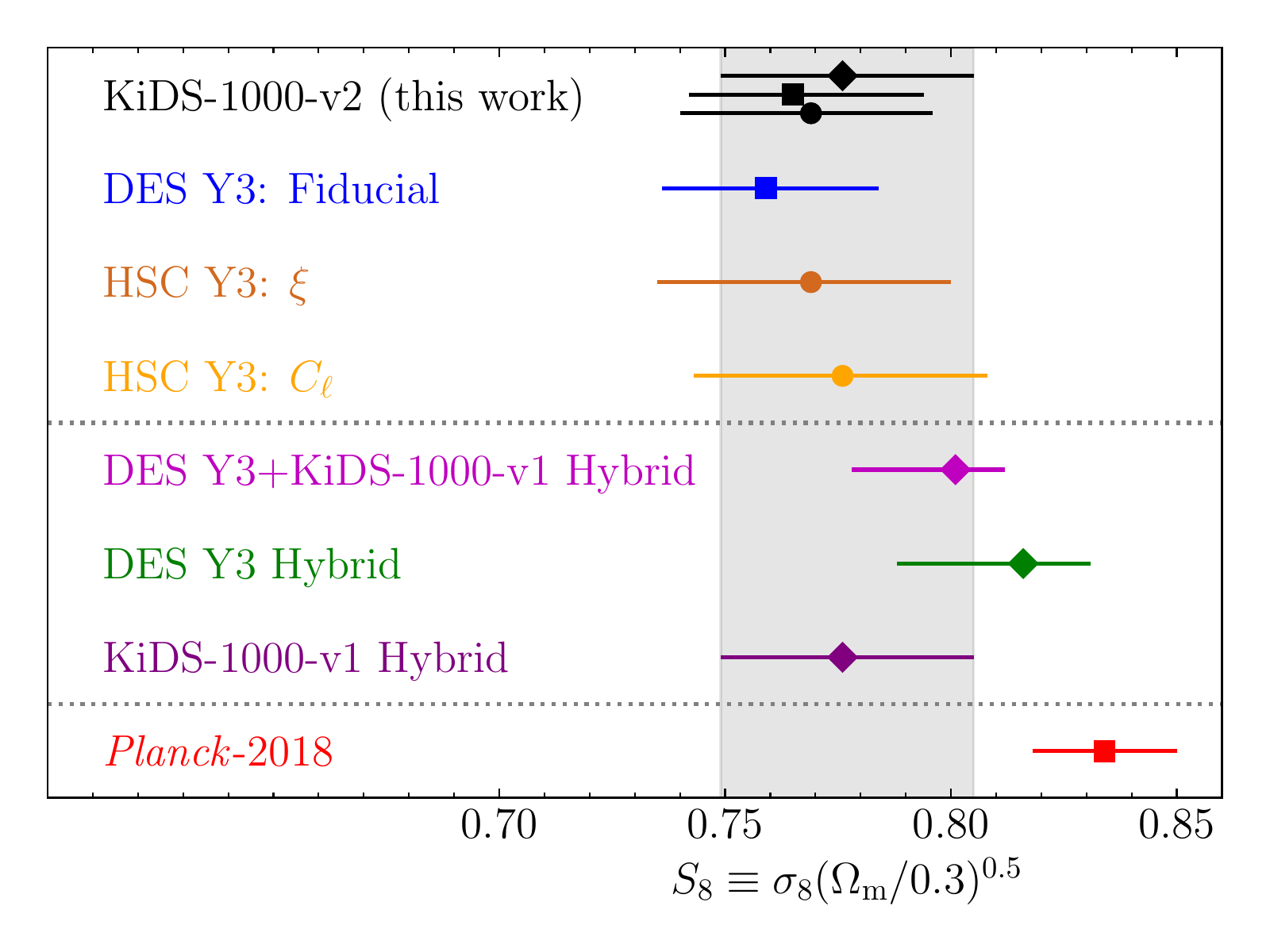}
      \caption{Marginalised constraints on $S_8$ derived from our fiducial analysis with \textsc{PolyChord}, compared with those from other contemporary cosmic shear surveys and the \textit{Planck} CMB analysis. Three sections, separated by dotted horizontal lines, indicate results of different origins. The first section includes results from individual cosmic shear surveys with their own analysis pipelines. The second section presents results from a collaborative effort between the DES and KiDS teams, who built a hybrid pipeline for analysing the data from both groups (\DK). The final section displays results from the \textit{Planck} CMB analysis. Different labels are used for different statistical methods: the diamond represents results using the MAP and PJ-HPD statistics, the square denotes the mean-marginal statistics, and the circle shows the maximum-marginal statistics. The error bars correspond to the $68\%$ credible intervals.} 
         \label{fig:S8Fiducial}
  \end{figure}

Figure~\ref{fig:S8Fiducial} presents our primary $S_8$ constraints and compares them with those from other contemporary cosmic shear surveys and the \textit{Planck} CMB analysis. For ease of comparison, we show all three summary statistics for our fiducial results, while for other surveys, we display their headline values, as per their preferred summary statistics. Overall, our results align well with those from all major contemporary cosmic shear surveys. 

We note that our fiducial analysis pipeline is similar to the {\DK} Hybrid pipeline with one notable difference: while {\DK} included a free neutrino parameter, we kept the total neutrino mass fixed. {\DK} showed that this additional degree of freedom in the cosmological parameter space can slightly increase the projected marginal $S_8$ values relative to an analysis with a fixed neutrino mass. However, since we refer to their MAP and PJ-HPD results in Fig.~\ref{fig:S8Fiducial}, the comparison should not be influenced by these projection effects (for more details, refer to the discussion in {\DK}).

It is interesting to note that our fiducial results align almost identically with the KiDS-1000-v1 re-analysis conducted by {\DK}, who used the {\Aone} redshift calibration. This alignment arises from a balance of several effects in our analysis. Our improved shear calibration tends to increase $S_8$, while the enhanced {\Bone} redshift calibration tends to lower it. Moreover, thanks to our enhanced empirical corrections for PSF leakages, our $S_8$ constraints are less affected by changes in the small scale cut used in measuring two-point correlation functions. The shifts we observe are roughly two times smaller than those in the KiDS-1000-v1 re-analysis conducted by {\DK}, which we discuss in more detail in Sect.~\ref{Sec:scaleCut}. This helps reconcile the minor difference between our results and those of {\Aone}. 

\subsection{Impact of shear biases}
\label{Sec:resShear}

The primary aims of this study are to assess the impact of higher-order shear biases on the final parameter constraints and to develop a methodology for effectively addressing shear calibration uncertainties. Both of these aims can be achieved by examining the shifts in the constrained cosmological parameters resulting from different shear bias scenarios. As discussed in Sect.~\ref{Sec:analysis} and Appendix~\ref{Sec:Dm}, the residual shear biases have only a minor effect on the measured data vector. This allows us to determine the shifts in the best-fit values of the constrained parameters using a local minimisation algorithm, such as the Nelder-Mead method~\citep{Nelder10.1093/comjnl/7.4.308}. These shifts in the best-fit values indicate the additional uncertainties stemming from systematic uncertainties in shear calibration.

\begin{figure}
  \centering
  \includegraphics[width=\hsize]{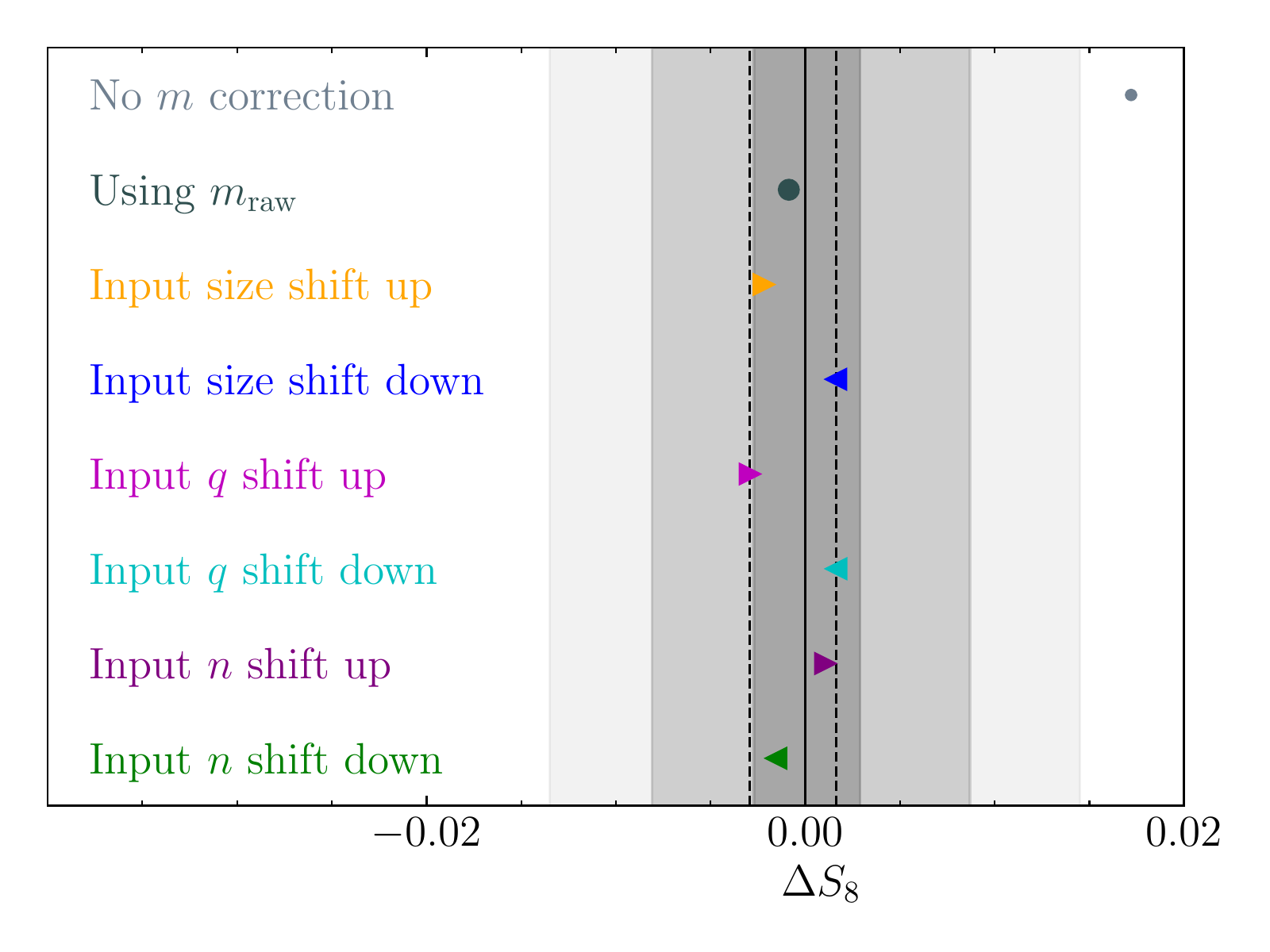}
      \caption{Shifts in best-fit values of $S_8$ under different residual shear bias scenarios. The shift, $\Delta S_8$, is calculated as $\Delta S_8=S_{8}^{\rm test} - S_{8}^{\rm fiducial}$, where $S_{8}^{\rm test}$ represents the best-fit values in the test scenarios determined by a local minimisation method that uses the best-fit values from the fiducial analysis ($S_{8}^{\rm fiducial}$) as a starting point. The grey shaded regions represent different percentiles of the credible intervals derived from our fiducial \textsc{PolyChord} run. From the innermost to the outermost region, these percentiles are $6.8\%$, $20.4\%$, and $34\%$, corresponding to $0.1$, $0.3$, and $0.5$ fractions of the reported sampling uncertainties. The dashed lines display the maximum shifts encountered in the six sets of morphology test simulations. These maximum shifts are used as the additional uncertainties in the reported best-fit values to account for the systematic uncertainties arising from shear calibration.} 
         \label{fig:S8shift}
  \end{figure}

Figure~\ref{fig:S8shift} shows shifts in our primary $S_8$ constraints for different residual shear bias scenarios. For comparison, we also include a shaded region denoting different levels of PJ-HPD credible intervals, as derived from our fiducial \textsc{PolyChord} chain. Apart from the extreme case where no shear calibration is applied, all other residual shear bias scenarios result in shifts less than 10 per cent of the initial sampling uncertainties. Notably, neglecting the higher-order correction for the shear-interplay effect and uncertainties in PSF modelling results in a negligible shift of only $-0.03\sigma$ (labelled `Using $m_{\rm raw}$' in the figure). This finding reinforces the reliability of previous KiDS cosmic shear analyses, which did not consider these higher-order effects.

The $S_8$ shifts, resulting from the input morphology test simulations, indicate additional systematic uncertainties in our shear bias calibration. To generate these test simulations, we changed the input values for three morphological parameters of the adopted S\'ersic profile: the half-light radius (labelled `size' in the figure), axis ratio (labelled `$q$'), and the S\'ersic index (labelled `$n$'). The adjustments were based on the fitting uncertainties reported by \citet{Griffith2012ApJS}, from whose catalogue we derived the input morphology for our simulations (refer to Sect.~2.1.2 of {\Litwothree}). For simplicity, we shifted all galaxies in the same direction for each test simulation, implicitly assuming that the fitting uncertainties stem from a coherent bias in that direction. This means that our test results represent the most extreme scenario. To consider both directions, we adjusted the input values in both positive and negative directions, leading to a total of six test simulations. Further details regarding the generation and comparison of these test simulations can be found in Appendix~\ref{Sec:Dm}.

We observe that shifts in the input galaxy axis ratio lead to the most significant changes in $S_8$: a $-0.10\sigma$ shift for increased input axis ratio and a $+0.06\sigma$ shift for decreased input axis ratio. This behaviour aligns with our expectations for the \textit{lens}fit code employed in our analysis. As it incorporates prior information on measured galaxy ellipticities during its Bayesian fitting process, it is more sensitive to changes in the distributions of sample ellipticities.

These $S_8$ shifts, obtained from the test simulations, provide a quantitative measure of the potential impact of inaccuracies in the input morphology and the sensitivity of the \textit{lens}fit code to the underlying sample morphology distributions. When presenting the $S_8$ constraints, we accounted for these systematic uncertainties by including the maximum shifts in the reported uncertainties. In other words, we considered the shifts corresponding to the changes in input axis ratio (represented as dashed lines in Fig.~\ref{fig:S8shift}), from the six sets of test simulations, as additional systematic uncertainties. These are reported alongside the original statistical uncertainties from the main sampling chain. It should be noted that these additional systematic uncertainties are specific to the SKiLLS image simulations and the \textit{lens}fit shape measurement code used in our analysis. To reduce these uncertainties, future advancements in shear measurements should focus on improving the realism of image simulations and enhancing the robustness of the shear measurement algorithm.

\subsection{Impact of altering inference setups}
\label{Sec:resPrevious}

Although our main updates revolve around the shear measurement and calibration, we also implemented several modifications to the cosmological inference pipeline, drawing upon recent developments from {\DK}. As such, it is beneficial to conduct some extended runs with various setup configurations. 

For these test runs, we employed \textsc{MultiNest} as our sampling code, as it operates approximately five times faster than \textsc{PolyChord}, but at the cost of underestimating the width of the posterior distributions and thus the reported uncertainties by about $10\%$. However, the best-fit values from \textsc{MultiNest} are not biased (as evident in Fig.~\ref{fig:S8OmFiducial}). Thus, comparisons made using \textsc{MultiNest} will yield conservative but unbiased results.

\begin{table*}
    \centering
    \caption{Point estimates for $S_8$ from different inference setups.}
    \label{tab:test}
    \renewcommand{\arraystretch}{1.5}
    \begin{tabular}{l|ccc|cc|cc}
        \hline\hline
        \multicolumn{1}{l|}{Setups} & \multicolumn{3}{c|}{MAP \& PJ-HPD} & \multicolumn{2}{c|}{Mean marginal} & \multicolumn{2}{c}{Maximum marginal} \\
        \hline
         & $\chi^2$ & $S_8$ & $\Delta S_8$ & $S_8$ & $\Delta S_8$ & $S_8$ & $\Delta S_8$ \\
         \hline
         Fiducial: \textsc{PolyChord} & $62.7$ & $0.776_{-0.027-0.003}^{+0.029+0.002}$ & - & $0.765_{-0.023}^{+0.029}$ & - & $0.769_{-0.029}^{+0.027}$ & - \\
         \hline
         Fiducial: \textsc{MultiNest} & $62.7$ & $0.774_{-0.023}^{+0.027\hphantom{+0.002}}$ & $\hphantom{-}0.00\sigma$ & $0.765_{-0.017}^{+0.028}$ & $\hphantom{-}0.00\sigma$ & $0.771_{-0.027}^{+0.023}$ & $\hphantom{-}0.00\sigma$ \\
         \hline
         NLA: $A_{\rm IA}:\ [-6, 6]$ & $62.7$ & $0.774_{-0.020}^{+0.034\hphantom{+0.002}}$ & $\hphantom{-}0.02\sigma$ & $0.766_{-0.020}^{+0.028}$ & $\hphantom{-}0.03\sigma$ & $0.770_{-0.026}^{+0.023}$ & $-0.01\sigma$ \\
         NLA-z: $A_{\rm IA}, \eta_{\rm IA}:\ [-5, 5]$ & $62.7$ & $0.776_{-0.025}^{+0.022\hphantom{+0.002}}$ & $\hphantom{-}0.10\sigma$ & $0.765_{-0.020}^{+0.027}$ & $\hphantom{-}0.00\sigma$ & $0.770_{-0.025}^{+0.023}$ & $-0.05\sigma$ \\
         Scales: $0\farcm 5<\theta<300\arcmin$ & $64.5$ & $0.769_{-0.036}^{+0.014\hphantom{+0.002}}$ & $-0.17\sigma$ & $0.755_{-0.022}^{+0.029}$ & $-0.40\sigma$ & $0.763_{-0.030}^{+0.022}$ & $-0.31\sigma$ \\
        KiDS-1000-v1 setups & $64.9$ & $0.771_{-0.029}^{+0.017\hphantom{+0.002}}$ & $-0.13\sigma$ & $0.755_{-0.020}^{+0.026}$ & $-0.44\sigma$ & $0.759_{-0.024}^{+0.022}$ & $-0.50\sigma$ \\
        \hline
    \end{tabular}
    \renewcommand{\arraystretch}{1}
    \tablefoot{`Fiducial: \textsc{PolyChord}' denotes our headline results, which are the same as those presented in Table~\ref{tab:res}. `Fiducial: \textsc{MultiNest}' represents the same parameter setup as our fiducial analysis, but employs the \textsc{MultiNest} sampling code. We used this as the reference to assess test results because all test runs utilise the \textsc{MultiNest} code for increased speed. When comparing the Fiducial: \textsc{MultiNest} results with the primary \textsc{PolyChord} results, we can conclude that the best-fit values obtained from \textsc{MultiNest} are unbiased. The relative shift in $S_8$, denoted as $\Delta S_8$, is calculated by comparing the best-fit values from the test runs to the reference Fiducial: \textsc{MultiNest} run. The $\Delta S_8$ values are expressed as a fraction of $\sigma$, which signifies the standard deviation of estimates from the test run. We calculated $\Delta S_8$ for different summary statistics separately for consistency. For MAP \& PJ-HPD results, we also present the best-fit $\chi^2$ values. For comparison, the best-fit $\chi^2$ values from {\Aone} and {\Bone} are $82.2$ and $63.2$, respectively.}
\end{table*}

\subsubsection{Priors for the NLA model}

We began by testing the prior for the NLA model. As discussed in Sect.~\ref{Sec:analysis}, our fiducial analysis implemented a redshift-independent NLA model with a narrow flat prior for the amplitude parameter $A_{\rm IA}$. This model, motivated by the work of \citet{Fortuna2021MNRAS.501.2983F}, serves as an alternative to the uninformative broad prior previously used. To investigate the impact of this change on our final results, we performed two additional runs: one employing a redshift-independent NLA model with a broad $A_{\rm IA}$ prior ranging from $[-6, 6]$, in line with KiDS-1000-v1 analyses, and another allowing for a redshift-dependent IA amplitude, namely, the NLA-$z$ variant. The redshift evolution is modelled using a power law of the form $[(1+z)/(1+0.62)]^{\eta_{\rm IA}}$, with priors of $[-5, 5]$ for both $A_{\rm IA}$ and $\eta_{\rm IA}$, in line with {\DK}.

\begin{figure}
  \centering
  \includegraphics[width=\hsize]{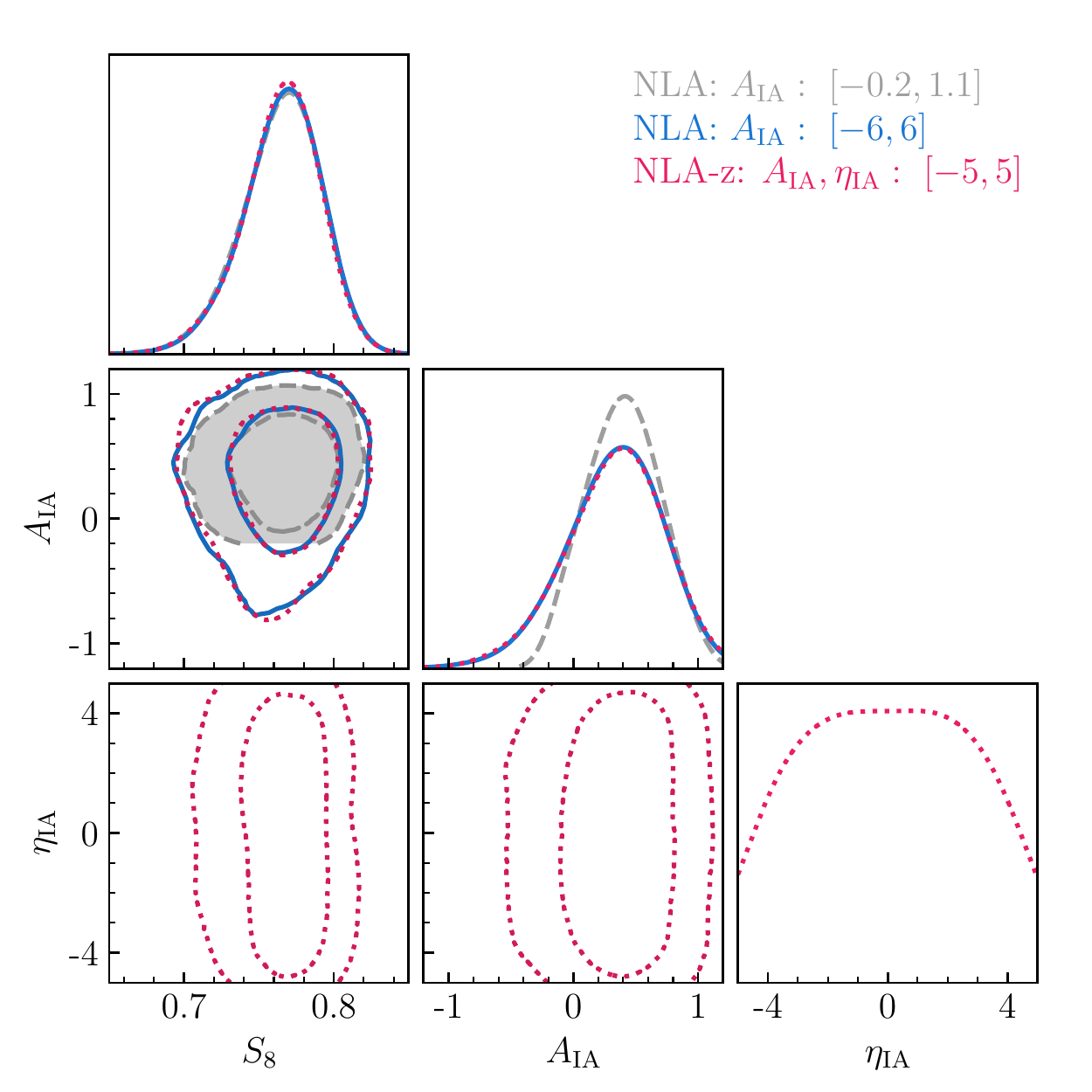}
      \caption{Comparison of projected posterior distributions for the parameters $S_8$, $A_{\rm IA}$, and $\eta_{\rm IA}$, derived from three different NLA prior setups. The contours correspond to the $68\%$ and $95\%$ credible intervals and are smoothed using Gaussian KDE with a bandwidth scaled by a factor of 1.5.} 
         \label{fig:S8IA}
  \end{figure}

Figure~\ref{fig:S8IA} presents a comparison of the posterior distributions obtained from the different NLA prior setups, and Table~\ref{tab:test} lists the point estimates for the critical $S_8$ parameter. We see consistent constraints on $S_8$ across all setups. The constrained $A_{\rm IA}$ under our narrower prior setup also aligns with those from the broad priors, albeit spanning a narrower range due to the constrained prior range, validating the prior range used in our fiducial analysis. Additionally, we observe that the $\eta_{\rm IA}$ parameter is not constrained by the data, suggesting that the use of the NLA-$z$ model may not be necessary for current weak lensing analyses.

\subsubsection{Different scale cuts}
\label{Sec:scaleCut}

In our fiducial analysis, we adopted a scale cut for the measured data vectors, ranging from $2\arcmin$ to $300\arcmin$, as suggested by {\DK}. This is a change from the KiDS-1000-v1 analyses, which used a range of $0\farcm 5<\theta<300\arcmin$. A re-analysis of KiDS-1000-v1 with this new scale cut by {\DK} led to a $0.7-0.8\sigma$ increase in the $S_8$ constraint. Using mock
analyses, they found that this offset could arise from noise fluctuations $23\%$ of the time.

\begin{figure}
  \centering
  \includegraphics[width=\hsize]{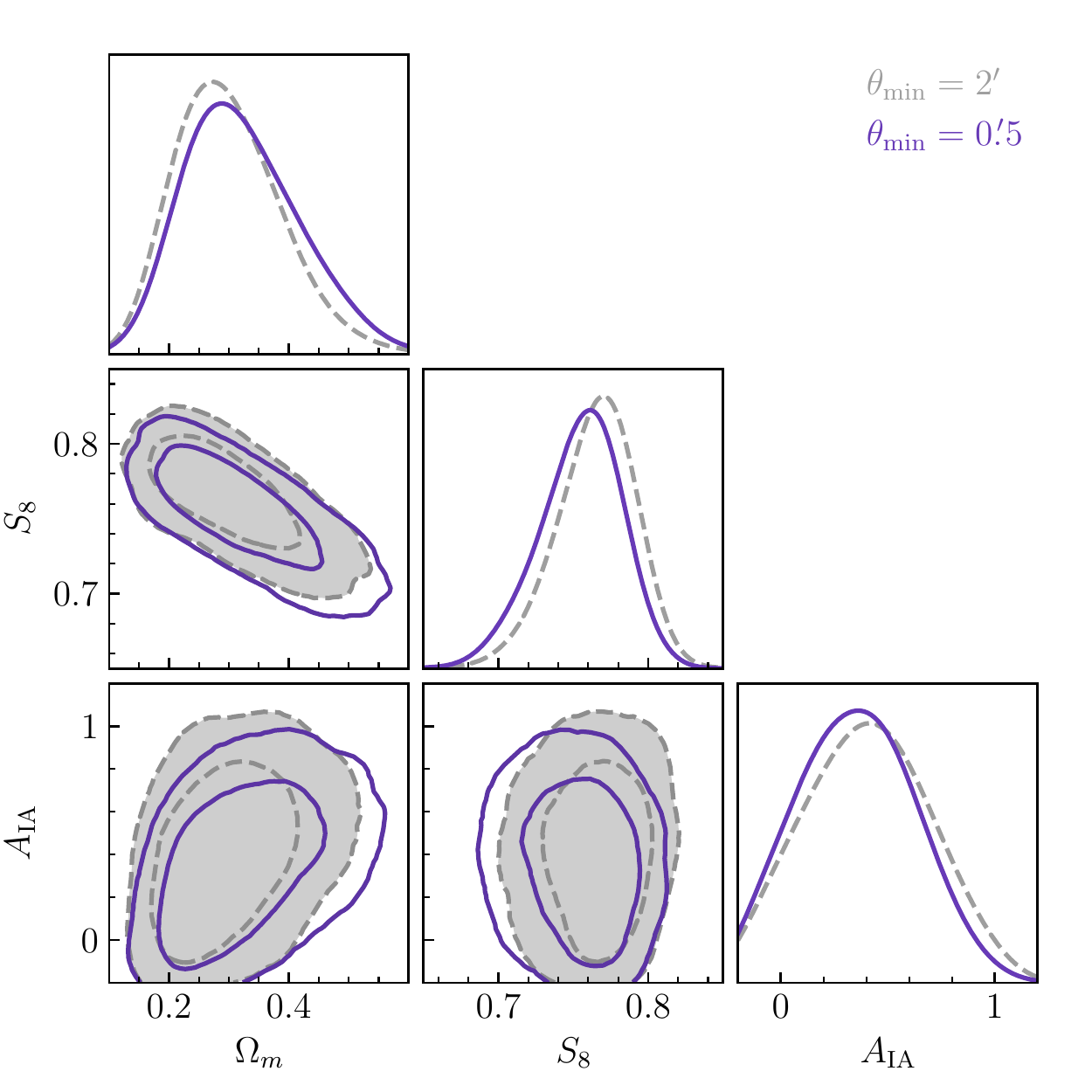}
      \caption{Comparison of projected posterior distributions for the parameters $\Omega_{\rm m}$, $S_8$, and $A_{\rm IA}$, derived from two different scale cuts. The contours correspond to the $68\%$ and $95\%$ credible intervals and are smoothed using Gaussian KDE with a bandwidth scaled by a factor of 1.5.} 
         \label{fig:S8scaleCut}
  \end{figure}

In light of the updates to our shear measurement, we revisited this test. Interestingly, as shown in Fig.~\ref{fig:S8scaleCut}, we observe a smaller difference between the two scale cuts than what was reported by {\DK}. Specifically, we observe shifts of $-0.17\sigma$, $-0.40\sigma$, and $-0.31\sigma$, corresponding to the MAP \& PJ-HPD, mean marginal, and maximum marginal summary statistics, respectively (refer to Table~\ref{tab:test} for exact values).

We attribute this increased robustness against small scale fluctuations to our improved empirical corrections of the PSF leakages into shear measurement. This is supported by Figs.~\ref{fig:Csys} and \ref{fig:Bmode}, where we see that the shear signals measured from the KiDS-1000-v2 catalogues exhibit overall smaller systematic errors. We note that \citet{Giblin2021AA...645A.105G} performed a mock test using the two-point correlation function and identified a change of less than $0.1\sigma$ in the $S_8$ constraints when the detected PSF residuals were incorporated into the KiDS-1000-v1 mock data. Nevertheless, it is plausible that these systematic effects have a more significant influence on COSEBIs, given their use of more sophisticated weighting functions~\citep{Schneider2010AA...520A.116S}. To quantify the improvements brought about by the updated shear measurements regarding the robustness of the COSEBIs, a similar mock analysis based on the COSEBIs statistic is warranted. We consider this an important topic for future study. For the current analysis, the test results simply affirm the robustness of our primary $S_8$ constraints.

\subsubsection{KiDS-1000-v1 setups}

\begin{figure}
  \centering
  \includegraphics[width=\hsize]{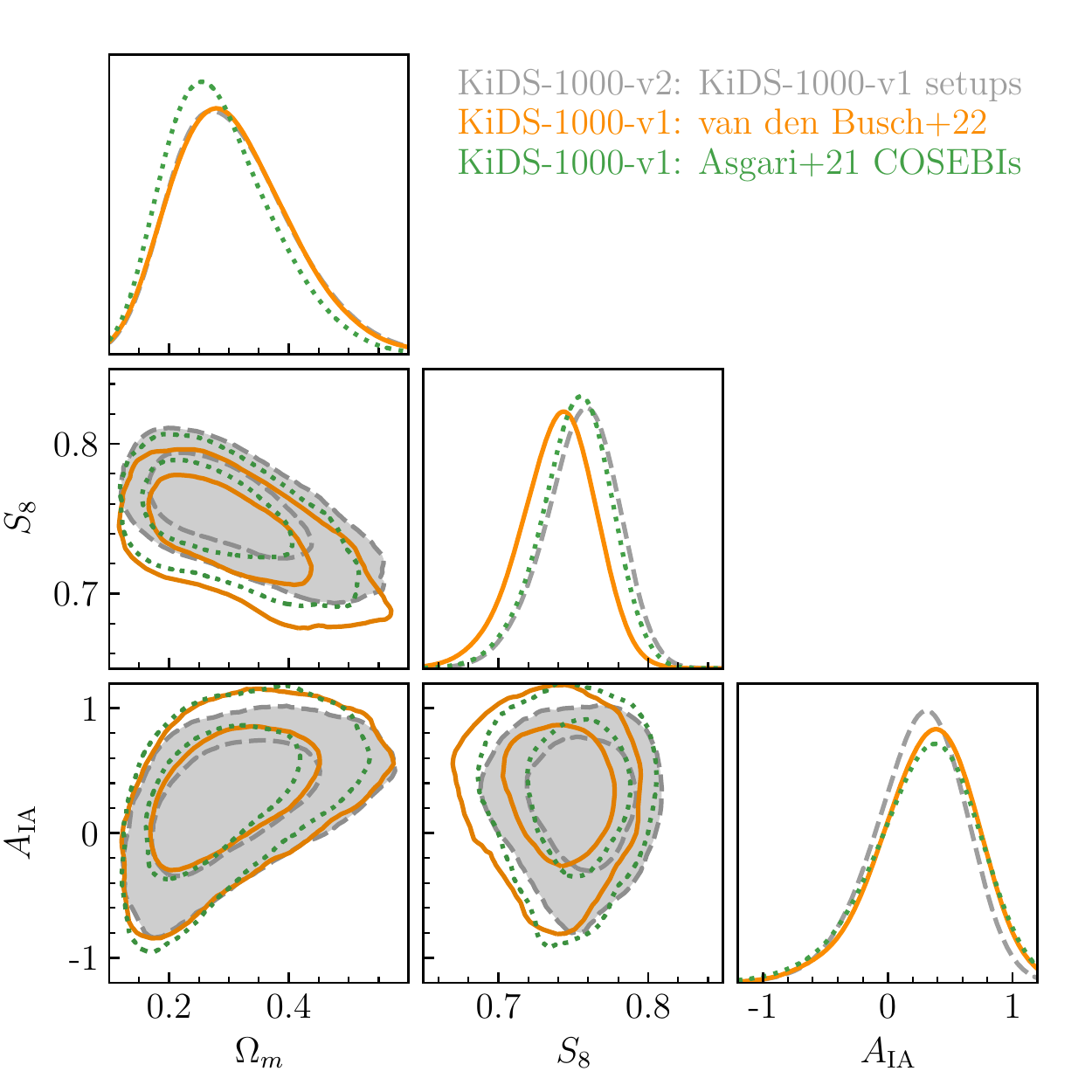}
      \caption{Comparison of projected posterior distributions for parameters $\Omega_{\rm m}$, $S_8$, and $A_{\rm IA}$ from our analysis (dashed grey lines) based on the KiDS-1000-v2 catalogue, to those from {\Bone} (solid orange lines) and {\Aone} (dotted green lines), both of which are based on the KiDS-1000-v1 catalogues. The cosmological inference pipeline and parameter priors are identical across all three analyses presented here. In terms of measurements, {\Bone} and {\Aone} used the same shear measurements and calibration, while {\Bone} and our analysis share the same redshift calibration. The contours correspond to the $68\%$ and $95\%$ credible intervals and are smoothed using Gaussian KDE with a bandwidth scaled by a factor of 1.5.} 
         \label{fig:S8KiDS1000}
  \end{figure}

To draw a direct comparison with the KiDS-1000-v1 results and evaluate the impact of our improved shear measurements and calibration, we performed a test run using the same inference pipeline and parameter priors as in the KiDS-1000-v1 analyses conducted by {\Aone} and {\Bone}. The differences compared to our fiducial analysis setup include: measurements from scales of $0\farcm 5$ to $300\arcmin$, use of the older version of \textsc{hmcode}, sampling with the \textsc{MultiNest} code, and a broad $A_{\rm IA}$ prior ranging from $[-6, 6]$. As shown in Fig.~\ref{fig:S8KiDS1000}, our test results are well aligned with the outcomes of the analyses by {\Aone} and {\Bone}. Notably, our new results show an increase in the $S_8$ value relative to {\Bone}, bringing it closer to the result obtained by {\Aone}.

We re-emphasise that our redshift calibration aligns with that of {\Bone}, who expanded the redshift calibration sample to more than double the size used by {\Aone} (see Appendix~\ref{Sec:zCalib} for details). This means that our redshift-related selection function closely mirrors that used in the {\Bone} sample. However, due to changes in the weighting and selection scheme between the KiDS-1000-v2 catalogue and the KiDS-1000-v1 catalogue, our sample cannot be considered as directly comparable to theirs. 

To provide a more quantitative understanding of the sample differences among the three analyses, we compared the effective number density of the source sample in our analysis to those used in {\Aone} and {\Bone}. The differences for each tomographic bin are $9.6\%$, $9.8\%$, $6.1\%$, $10.6\%$, and $2.8\%$ when compared to {\Aone}; and $-1.8\%$, $-1.3\%$, $-0.7\%$, $0.7\%$, and $3\%$ when compared to {\Bone}. Here, positive values signify an increase, while negative values denote a decrease. The differences between our catalogue and that of {\Aone} stem from both shear measurement and redshift calibration, whereas the difference between ours and that of {\Bone} arises mainly from the shear measurement, as we used the same SOM for the `gold' selection (see Appendix~\ref{Sec:zCalib}). As such, comparing our results directly with those of {\Bone} can provide clearer insights into the impact of our improvements in shear measurements. It is also worth noting that the increased effective number density in high redshift bins compared to {\Bone} is largely due to the increased weighting of faint objects in the updated version of \textit{lens}fit code. However, this comes at the cost of increased sample ellipticity dispersion, with a maximum increase of $6\%$ found in the fifth bin. These subtle differences in the source catalogues change the noise properties of the samples. Consequently, even with perfect calibration in each study, we would not expect to derive identical cosmological constraints from each analysis.

Interestingly, the increase in number density from {\Aone} to {\Bone}, and in this work, does not significantly reduce the marginalised uncertainties of the final cosmological parameters. This can be largely attributed to the fact that the majority of the constraining power in the KiDS analysis comes from the high redshift bins, as illustrated in Fig.~7 of {\Aone}, whereas our increase in number density is most pronounced in the lower redshift bins. Additionally, changes in the redshift distributions, due to alterations in the redshift calibration sample, could further impact the final constrained uncertainties, as demonstrated in Table 3 of {\Bone}. Lastly, due to the intricate degeneracy among nuisance and cosmological parameters, caution should be used when inferring that an increased number density will directly lead to a reduction in the marginalised uncertainties of specific parameters.

\section{Summary}
\label{Sec:sum}

We have conducted a cosmic shear analysis using the KiDS-1000-v2 catalogue, which is an updated version of the public KiDS-1000(-v1) catalogue with respect to shear measurements and calibration. Under the assumption of a spatially flat $\Lambda$CDM cosmological model, we derived constraints on $S_8=0.776_{-0.027-0.003}^{+0.029+0.002}$ based on the MAP and PJ-HPD summary statistics. The second set of uncertainties was incorporated to account for the systematic uncertainties within our shear calibration. The mean-marginal and maximum-marginal values obtained from the same sampling chain are $0.765_{-0.023}^{+0.029}$ and $0.769_{-0.029}^{+0.027}$, respectively. Our results are consistent with earlier results from KiDS-1000-v1 and other contemporary weak lensing surveys but show a ${\sim}2.3\sigma$ level of tension with the \textit{Planck} CMB constraints.

The main improvements in our analysis, relative to the KiDS-1000-v1 cosmic shear analyses, are attributed to the enhanced cosmic shear measurement and calibration. These enhancements were achieved through the updated version of the \textit{lens}fit shape measurement code, a new empirical correction scheme for PSF contamination, and the newly developed SKiLLS multi-band image simulations, as detailed in {\Litwothree}. We verified the reliability of the new measurement via a series of catalogue-level null tests proposed by \citet{Giblin2021AA...645A.105G}. The results indicate that the KiDS-1000-v2 catalogue shows overall better control over measurement systematics compared to the KiDS-1000-v1 catalogues. This improvement in reducing measurement systematics helps in reducing noise in small scale measurements, thereby enhancing the robustness of our cosmological parameter constraints against varying scale cut choices.

Our methodology for shear calibration largely aligns with the one detailed in {\Litwothree}, where we accounted for higher-order blending effects that arise when galaxies from different redshifts are blended, as well as the uncertainties in PSF modelling. However, when comparing the outcomes from the shear calibration with and without these higher-order adjustments, we find that these effects have a negligible impact on the present weak lensing analysis, a conclusion that is in line with the findings of \citet{Amon2022PhRvD.105b3514A}.

We recommend treating the statistical and systematic uncertainties from the shear calibration separately, given their distinct origins. The statistical uncertainties, which are determined by the simulation volume, can be reduced and incorporated into the covariance matrix used for cosmological inference. On the other hand, systematic uncertainties, associated with the realism of image simulations and the sensitivity of the shape measurement algorithm, can be more effectively addressed when considered as residual shear biases post-calibration. Assuming these residual shear biases are small, a forward modelling approach, combined with a local minimisation method, can be used to estimate their impact on the final parameter constraints. In our analysis, these additional systematic uncertainties contribute roughly $8\%$ of the final uncertainty on $S_8$. However, ongoing efforts to enhance shear measurement and calibration, such as increasing the realism of image simulations through Monte Carlo control loops~\citep{Refregier2014PDU.....3....1R} and leveraging new techniques such as \textsc{Metacalibration/Metadetection}~\citep{Huff2017arXiv170202600H,Sheldon2017ApJ...841...24S,Sheldon2020ApJ...902..138S,Hoekstra2021AA...646A.124H} to improve measurement robustness against underlying sample properties, may well lead to a reduction in these additional systematic uncertainties.

In our fiducial analysis, we opted for a redshift-independent NLA model with a narrow flat prior for the IA amplitude parameter, $A_{\rm IA}$, motivated by the work of \citet{Fortuna2021MNRAS.501.2983F}. However, we also investigated two alternative scenarios: one with a broad $A_{\rm IA}$ prior for the redshift-independent NLA model, echoing the KiDS-1000-v1 analysis by {\Aone}, and the other the NLA-$z$ variant, allowing for redshift evolution of the IA amplitude, as per the recent joint DES Y3+KiDS-1000 cosmic shear analysis (\DK). In all three scenarios, we find fully consistent constraints for $S_8$ and $A_{\rm IA}$, which indicates that the impact of the variations is negligible in these scenarios. To better understand the IA signals and their impact on cosmic shear analyses, future tests need to implement more substantial variations in IA models, for instance the halo model formalism introduced by \citet{Fortuna2021MNRAS.501.2983F}. Such an exploration would not only enhance our understanding of the measured IA signals, but also help mitigate correlations between nuisance parameters, thereby improving the precision of future cosmic shear analyses.


\begin{acknowledgements}
    We acknowledge support from: the Netherlands Research School for Astronomy (SSL); the Netherlands Organisation for Scientific Research (NWO) under Vici grant 639.043.512 (HHo); the Royal Society and Imperial College (KK); the Polish National Science Center through grants no.\ 2020/38/E/ST9/00395, 2018/30/E/ST9/00698, 2018/31/G/ST9/03388 and 2020/39/B/ST9/03494 (MB); the Polish Ministry of Science and Higher Education through grant DIR/WK/2018/12 (MB); the Royal Society through an Enhancement Award (RGF/EA/181006) and the Royal Society of Edinburgh for support through the Saltire Early Career Fellowship with ref.\ number 1914 (BG); the European Research Council (ERC) under Grant number 647112 (CH) and Consolidator Grant number 770935 (HHi, JLvdB, AHW, RR); the Max Planck Society and the Alexander von Humboldt Foundation in the framework of the Max Planck-Humboldt Research Award endowed by the Federal Ministry of Education and Research (CH); the UK Science and Technology Facilities Council (STFC) under grant ST/V000594/1 (CH), ST/V000780/1 (BJ) and ST/N000919/1 (LM); Heisenberg grant of the Deutsche Forschungsgemeinschaft grant Hi 1495/5-1 (HHi); CMS-CSST-2021-A01 and CMS-CSST-2021-A04, NSFC of China under grant 11973070 (HYS); Key Research Program of Frontier Sciences, CAS, Grant No. ZDBS-LY-7013 (HYS); Program of Shanghai Academic/Technology Research Leader (HYS). The results in this paper are based on observations made with ESO Telescopes at the La Silla Paranal Observatory under programme IDs: 088.D-4013, 092.A-0176, 092.D-0370, 094.D-0417, 177.A-3016, 177.A3017, 177.A-3018 and 179.A-2004, and on data products produced by the KiDS consortium. The KiDS production team acknowledges support from: Deutsche Forschungsgemeinschaft, ERC, NOVA and NWO-M grants; Target; the University of Padova, and the University Federico II (Naples). Contributions to the data processing for VIKING were made by the VISTA Data Flow System at CASU, Cambridge and WFAU, Edinburgh. Author contributions: All authors contributed to the development and writing of this paper. The authorship list is given in three groups: the lead authors (SSL, HHo, KK) followed by two alphabetical groups. The first alphabetical group includes those who are key contributors to both the scientific analysis and the data products. The second group covers those who have either made a significant contribution to the data products, or to the scientific analysis.
\end{acknowledgements}

  \bibliographystyle{aa} 
  \bibliography{reference} 

\begin{appendix} 

\section{Shear bias in two-point statistics}
\label{Sec:CF4m}

When calibrating the shear measurements in the two-point correlation function, it is usually assumed that the correlations involving the shear bias can be ignored, which includes correlations between different tomographic and spatial angular bins. This simplification leads to the following relationship between the true correlation function of cosmic shear in tomographic bins $i,j$, denoted as $\xi^{ij}$, and the measured signal $\hat{\xi}^{ij}$:
\begin{equation}
\label{eq:mTwoPoint}
\begin{split}
\hat{\xi}^{ij}(\theta) &= \langle\hat{\gamma}^{i}(\theta')~\hat{\gamma}^{j}(\theta'+\theta)\rangle\\
& = \langle[1+m^{i}(\theta')]~[1+m^{j}(\theta'+\theta)]~\gamma^{i}(\theta')~\gamma^{j}(\theta'+\theta)\rangle\\
& = \langle[1+m^{i}(\theta')]~[1+m^{j}(\theta'+\theta)]\rangle~\xi^{ij}(\theta)\\
& \simeq (1+\overline{m^{i}})~(1+\overline{m^{j}})~\xi^{ij}(\theta)~,
\end{split}
\end{equation}
where $\overline{m^{i}}$ is estimated by averaging over all sources in a given tomographic bin $i$, and we use $\langle\cdot\rangle$ to denote the correlation function. We also assumed that the shear bias is independent of the underlying shear to simplify the equation. The result of Eq.~(\ref{eq:mTwoPoint}) allows us to average the multiplicative biases over all the galaxies in a given tomographic bin to mitigate the individual noisy bias estimation. 

However, in principle, the shear bias can be scale dependent due to spatial fluctuations in source density~(e.g.~\citealt{Samuroff2018MNRAS.475.4524S}). With SKiLLS, we can directly examine these correlations by measuring the shear bias in the two-point estimators. We measured the shear correlation function in the SKiLLS mock catalogue using Eq.~(\ref{eq:CF}). Since we know the true $\xi^{ij}_{+}(\theta)=\gamma_{\rm input}^2$ in simulations, where $\gamma_{\rm input}$ is the amplitude of the constant input shear, we can estimate the shear bias in the two-point correlation function directly by comparing the measured $\hat{\xi}^{ij}_{+}$ to the input $\xi^{ij}_{+}$ following Eq.~(\ref{eq:mTwoPoint}). 

\begin{figure*}
   \centering
   \includegraphics[width=\textwidth]{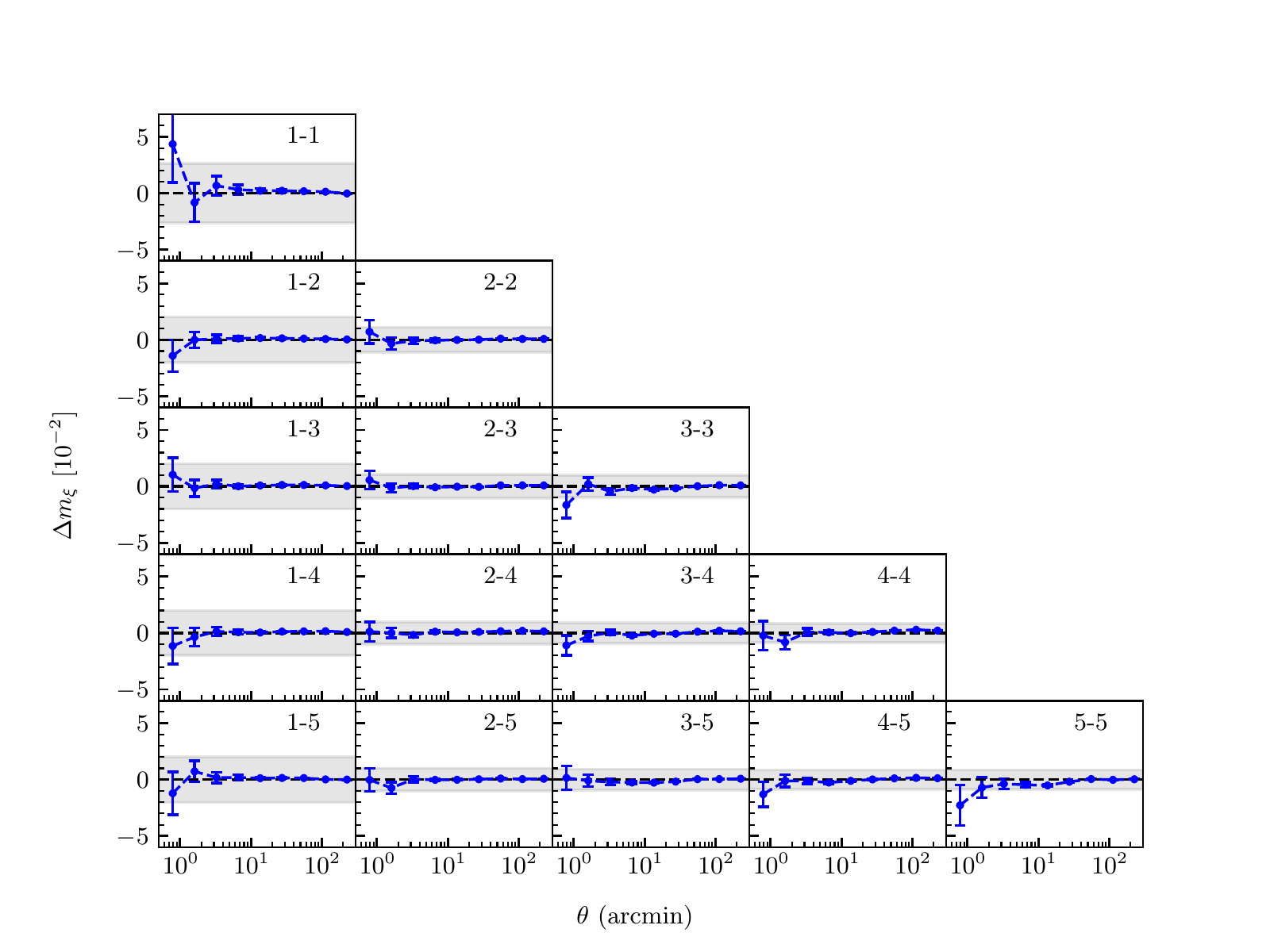}
   \caption{Two-point correlations between the multiplicative shear biases. The correlation is estimated as $\Delta m_{\xi} \equiv \hat{\xi}^{ij}_{+}/\gamma^{2}_{\rm input} - (1+\overline{m^{i}})~(1+\overline{m^{j}})$. The 15 panels represent the different combinations of the five redshift bins utilised in our cosmic shear analysis. The shaded regions within each panel denote the statistical uncertainties of our shear calibration for each tomographic bin, as outlined in Table~\ref{tab:mbias}.}
    \label{fig:CF4m}
    \end{figure*}

Figure~\ref{fig:CF4m} shows the difference between the shear biases with and without considering its correlations, defined as $\Delta m_{\xi} \equiv \langle[1+m^{i}(\theta')]~[1+m^{j}(\theta'+\theta)]\rangle - (1+\overline{m^{i}})~(1+\overline{m^{j}})$. It shows that the difference is negligible across all scales and tomographic bins, in agreement with the statistical uncertainties of our shear calibration, which are represented by the shaded regions. These findings confirm that we can neglect the correlations between shear biases in the current KiDS weak lensing analysis.

\section{SOM redshift calibration}
\label{Sec:zCalib}

This appendix provides information on the redshift calibration reference sample and SOM configurations used in our analysis. For a more comprehensive overview and validation of the SOM redshift calibration method in the KiDS analysis, we refer to \citet{Wright2020AA...640L..14W}, \citet{Hildebrandt2021AA...647A.124H}, and {\Bone}.  

We employed the fiducial spectroscopic sample described in {\Bone} as our calibration reference sample. This sample comprises spectroscopic redshift estimates (spec-$z$s) from various spectroscopic surveys that overlap with KiDS fields, enabling us to assign KiDS photometric measurements to objects in the reference sample. In cases where an object had multiple spectroscopic measurements, {\Bone} defined a specific hierarchy to select the most reliable redshift estimates based on the quality of the measurements. For further details on the adopted spectroscopic samples and the compilation procedure, readers are referred to Appendix A of {\Bone}. 

For our calibration, we used a $101\times101$ hexagonal SOM trained on the $r$-band magnitude and $36$ colours derived from the PSF-matched, list-driven nine-band $ugriZYJHK_s$ photometry from the KiDS+VIKING surveys. This SOM is identical to the fiducial SOM constructed in {\Bone}. We segregated the reference and target samples into the trained SOM cells separately for each tomographic bin, allowing us to create comparable groupings between the spectroscopic and photometric sources in each bin. During this process, we further categorised the original SOM cells using a hierarchical cluster analysis implemented by the `hclust' function within the \textsc{R} Stats Package\footnote{\url{https://www.rdocumentation.org/packages/stats/versions/3.6.2/topics/hclust}} to increase the number of galaxies per grouping. We adopted the same number of clusters per bin (4000, 2200, 2800, 4200, and 2000) as \citealt{Wright2020AA...640L..14W}, who determined these numbers using simulations produced by \citet{Busch2020AA...642A.200V}.

To mitigate the effects of photometric noise and the incompleteness of the reference sample, we applied an additional selection step to the SOM groupings. We excluded any grouping where the mean spectroscopic redshift of the reference sample $\overline{z_{\rm spec}}$ and the mean photometric redshift of the target sample $\overline{z_{\rm B}}$ exhibited a significant discrepancy, defined as $|\overline{z_{\rm spec}} - \overline{z_{\rm B}}| > 5\sigma_{\rm mad}$. Here, $\sigma_{\rm mad}$ represents the normalised median absolute deviation of all SOM groupings, which we calculated to be $0.122$ in our case. This step allowed us to define the KiDS `gold' sample, which we used to compute the redshift distributions and perform the cosmic shear analysis.

\section{Systematic uncertainties from the shear calibration}
\label{Sec:Dm}

In this appendix, we outline our approach to addressing the systematic uncertainties arising from shear calibration. Our methodology involves two primary steps: In Sect.~\ref{Sec:DmEsti}, we quantify the potential residual biases after implementing our simulation-based shear calibration. In Sect.~\ref{Sec:DmProp}, we propagate these systematic uncertainties into the final uncertainties of the estimated cosmological parameters.

We propose a separate accounting of the shear calibration uncertainties, as it is considered more accurate and informative than the traditional approach, which uses nominal shear calibration uncertainties that are deliberately overestimated to encompass potential systematic uncertainties arising from shear calibration. Our approach clearly illustrates the extent to which the final cosmological parameters of interest are influenced by these systematic uncertainties from shear calibration. 

Furthermore, as mentioned in Sect.~\ref{Sec:biasUncertainties}, these systematic uncertainties have fundamentally different origins from the statistical uncertainties incorporated in the covariance matrix. They represent the fundamental limitations of current simulation-based shear calibration methods. The limitations inherent in these systematic uncertainties cannot be eliminated by merely increasing the scale of image simulations. However they can be mitigated by empirically enhancing the realism of the image simulations, for example, using the Monte-Carlo Control Loop method~\citep{Refregier2014PDU.....3....1R}, or by improving the robustness of the shear measurement algorithm, such as the \textsc{Metacalibration/Metadetection} method~\citep{Huff2017arXiv170202600H,Sheldon2017ApJ...841...24S,Sheldon2020ApJ...902..138S,Hoekstra2021AA...646A.124H}.

\subsection{Quantifying residual shear biases with sensitivity analysis}
\label{Sec:DmEsti}

   \begin{figure*}
   \centering
   \includegraphics[width=\textwidth]{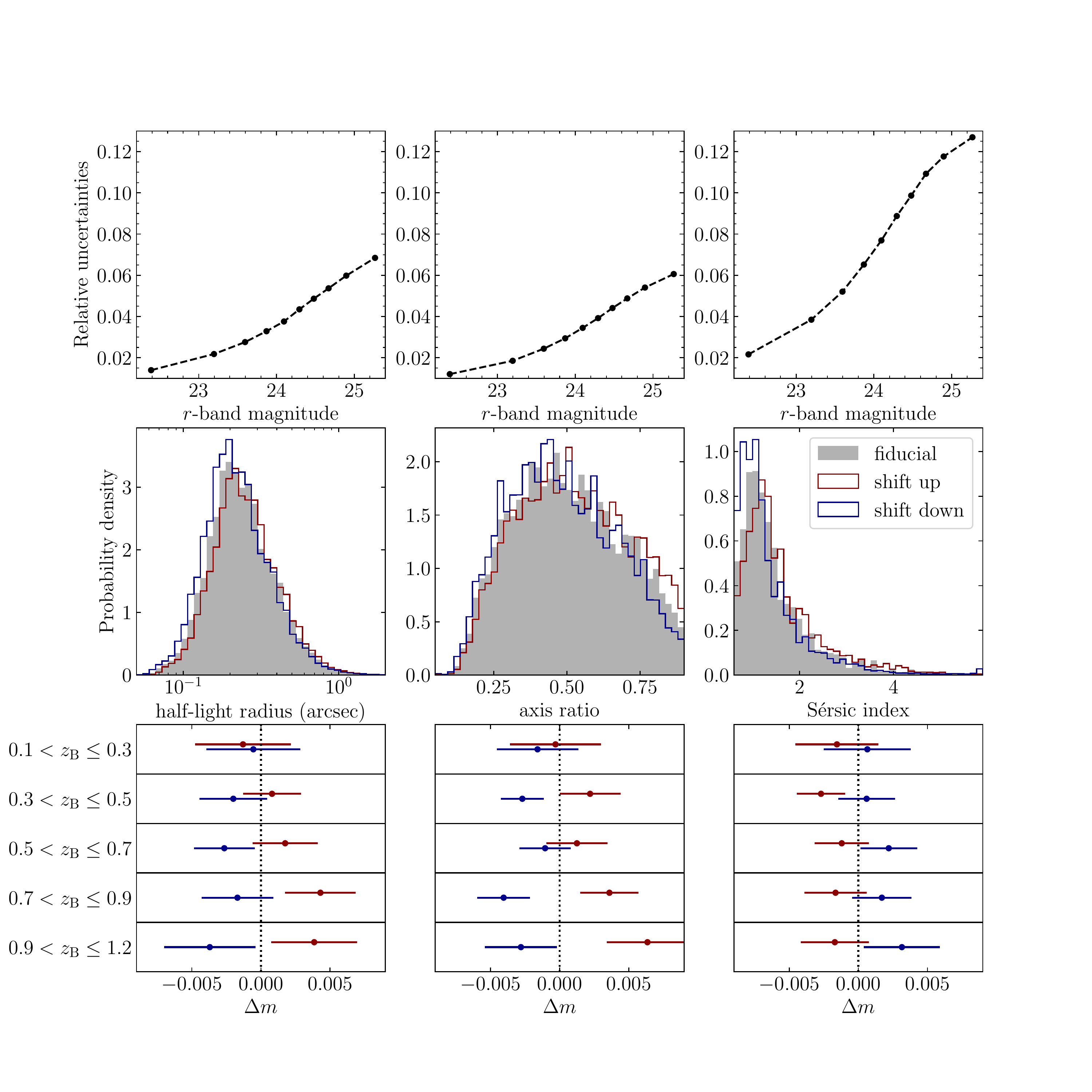}
   \caption{Changes in input morphological parameter values for assessing residual biases after simulation-based shear calibration. From left to right, the order is half-light radius, axis ratio, and S\'ersic index. \textit{Top panels}: Relationship between relative fitting uncertainties and the $r$-band magnitude, as measured from the catalogues of \citet{Griffith2012ApJS}. The values shown are calculated using the median values of the binned samples. \textit{Middle panels}: Overall distributions of input morphological parameters. Comparisons are drawn among the fiducial (grey shades), the test sample with input values increased by an amount corresponding to the relative uncertainties (red lines), and the test sample with input values decreased by the same amount of relative uncertainties (blue lines). \textit{Bottom panels}: Shifts in multiplicative biases in tomographic bins resulting from changes in input morphological parameter values. Both fiducial and test simulations are re-weighted using the same dataset via the method detailed in Sect.~\ref{Sec:calib2}.}
    \label{fig:dmMorp}
    \end{figure*}

Residual biases may persist after simulation-based shear calibration due to imperfect alignment between simulations and data, as elucidated by {\Kone}. These discrepancies pose challenges for shear calibration methods dependent on image simulations and underscore the need for re-weighting simulations to more closely align with the data. However, given that intrinsic galaxy properties in real data are unknown, this re-weighting process relies on noisy measured properties, rendering it vulnerable to calibration selection biases as discussed by {\FC}. The uncertainties linked with the measured properties cause galaxies to be intermixed among defined bins, leading to the up-weighting or down-weighting of certain galaxies. As a result, even if the re-weighted sample aligns with the data in terms of the distribution of measured properties, it does not ensure identicality in terms of intrinsic properties. In other words, shear biases can still vary between two samples with identical distributions of apparent measured properties. Our aim is to quantify these residual biases and incorporate them into the final uncertainties of cosmological parameters.

The SKiLLS multi-band image simulations used in this analysis incorporate several enhancements, informed by insights gathered from previous KiDS simulation studies~({\FC}; {\Kone}). These improvements include: reproducing variations in star density, PSF, and noise background across the KiDS footprint; incorporating faint galaxies down to an $r$-band magnitude of 27 to account for correlated noise from undetected objects~(e.g.~\citealt{Hoekstra2017MNRAS.468.3295H}); including realistic clustering from $N$-body simulations to address blending effects~(e.g.~{\Kone}); and adopting an end-to-end approach for photo-$z$ estimation to account for photo-$z$ measurement uncertainties. These improvements augment the robustness of the shear biases estimated from SKiLLS against various observational conditions.

In an investigation on the propagation of observational biases in shear surveys, \citet{Kitching2019OJAp....2E...5K} demonstrated that the measured shear power spectrum is, to first order, predominantly influenced by the mean of the multiplicative bias field across a survey. This suggests that if the shear bias estimated from simulations accurately reflects the mean value of the targeted sample, the shear calibration will be robust enough for KiDS-like cosmic shear analyses. Therefore, we conclude that potential residual biases related to observational conditions have negligible influence on our shear calibration, and we focused on systematic uncertainties arising from galaxy morphology uncertainties, specifically the assumed S\'ersic profile and its parameters derived from \textit{Hubble} Space Telescope observations~\citep{Griffith2012ApJS}. For a model-fitting shape measurement code like \textit{lens}fit, these galaxy morphology uncertainties are the main sources of residual shear biases after implementing the simulation-based shear calibration. 

The deviation from the S\'ersic profile is challenging to address for the current SKiLLS simulations, as our copula-based learning algorithm requires a parameterised model for its application. However, the S\'ersic model has been validated as sufficient for KiDS-like analyses by {\Kone}, who used the same morphology catalogue as our work. Thus, we focus on the measurement uncertainties of the S\'ersic parameters: half-light radius, axis ratio, and S\'ersic index. We first examined the fitting uncertainties reported by \citet{Griffith2012ApJS} to assess the accuracy of these parameters in our input catalogue. We find that the median relative uncertainties for these parameters are a smooth function of galaxy magnitude, as shown in the top panels of Fig.~\ref{fig:dmMorp}. This allows us to capture these correlations through simple linear interpolation. 

We interpreted these relative uncertainties as indicators of the systematic uncertainties in our input morphology. We assumed the most extreme scenarios, in which these measured statistical uncertainties are all caused by a coherent bias in the same direction. Consequently, we adjusted all galaxies in our input sample in the same direction, with the amplitude of the adjustment determined based on their $r$-band magnitude using a simple linear interpolation of the measured median correlations. We examined shifts towards both larger and smaller values and considered the three S\'ersic parameters separately. This resulted in six test simulations corresponding to the six different sets of variations in input morphology parameter values. The input parameter distributions for these test simulations, as shown in the middle panels of Fig.~\ref{fig:dmMorp}, are compared to the distributions of the fiducial simulations. A clear shift of the entire distribution is evident, suggesting that our test simulations represent the most extreme scenarios in which the measured statistical uncertainties are coherently biased in the same direction, a situation that is unlikely in reality. Therefore, the residual biases we identified from these test simulations provide a conservative estimate. 

We applied the same data analysis procedures to the test simulations as we did to the fiducial simulations, including shear and redshift estimates. We also followed the same re-weighting procedure for the test simulations as for the fiducial simulations, ensuring that the calibration selection biases are also captured. The differences in shear biases between these test simulations and our fiducial simulation are illustrated in the bottom panels of Fig.~\ref{fig:dmMorp}. The small differences indicate that the residual shear biases, after implementing our fiducial shear bias calibration, are insignificant.

\subsection{Propagating residual shear biases with forward modelling}
\label{Sec:DmProp}

Accurately incorporating the systematic uncertainties from shear calibration into the covariance matrix presents a challenge, as residual shear biases directly scale the data vector, as shown in Eq.~(\ref{eq:mTwoPoint}). A more direct approach is to assess the shift in the measured shear signal caused by the residual shear biases and evaluate how these data vector shifts influence the constrained cosmological parameters. Given the minor residual shear biases illustrated in Fig.~\ref{fig:dmMorp} and the unchanged covariance, it is not necessary to reiterate the sampling of the posterior distributions for each shift. Instead, we can implement a local minimisation algorithm to find nearby best-fit values for each shift, using starting points from the fiducial sampling chain. The range of these new best-fit values, each associated with a shift, indicates the additional systematic uncertainties introduced by the residual shear biases.

This approach naturally integrates with our existing cosmological inference method, as outlined in Sect.~\ref{Sec:analysis}, which already requires an additional local optimisation step to refine the best-fit values identified by the sampling code. We simply replicated this optimisation step, using the original best-fit value as the starting point and the shifted likelihood to determine the best-fit values associated with various alterations in measured signals. The variability in these test best-fit values provides an expanded credible region for the inferred parameters, thereby representing the systematic uncertainties from shear calibration. We included these additional uncertainties when presenting the point estimates of our primary parameters (see Sect.~\ref{Sec:resShear} for details).

\section{Contour plots for all free parameters}
\label{Sec:allPara}

In this appendix, we provide two supplementary contour plots that display the posterior distributions of all twelve free parameters from our fiducial analyses, as produced by both the \textsc{PolyChord} and \textsc{MultiNest} sampling codes. The overall concordance between the results generated by \textsc{PolyChord} and \textsc{MultiNest} is evident.

   \begin{figure*}
   \centering
   \includegraphics[width=\textwidth]{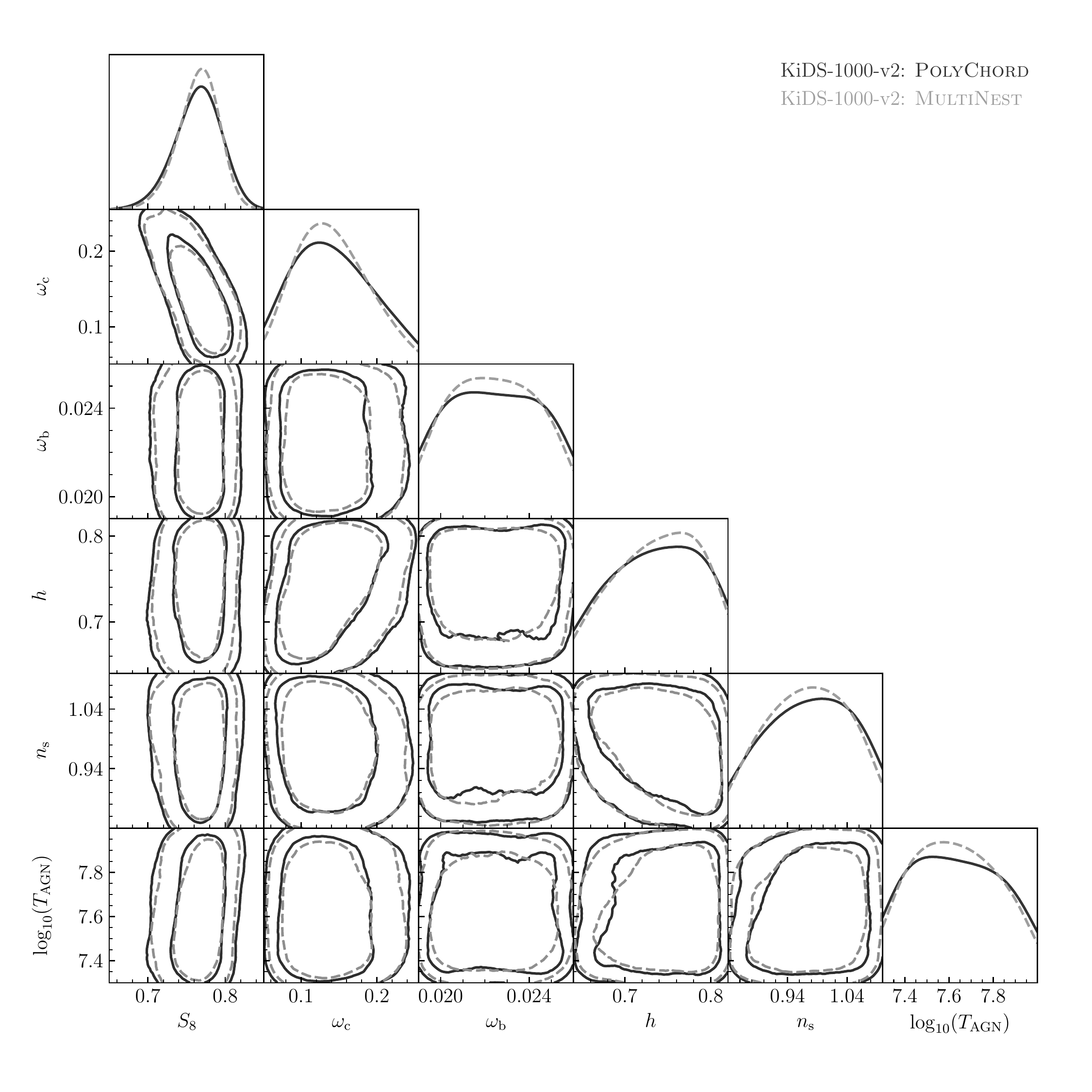}
   \caption{Posterior distributions of cosmological and baryonic parameters from our fiducial analyses, as generated by \textsc{PolyChord} (solid black lines) and \textsc{MultiNest} (dashed grey lines). The contours represent the $68\%$ and $95\%$ credible intervals and are smoothed using a Gaussian KDE with a bandwidth scaled by a factor of 1.5. We note that $S_8$ is the only parameter that our data robustly constrain.}
    \label{fig:S8cosmo}
    \end{figure*}

   \begin{figure*}
   \centering
   \includegraphics[width=\textwidth]{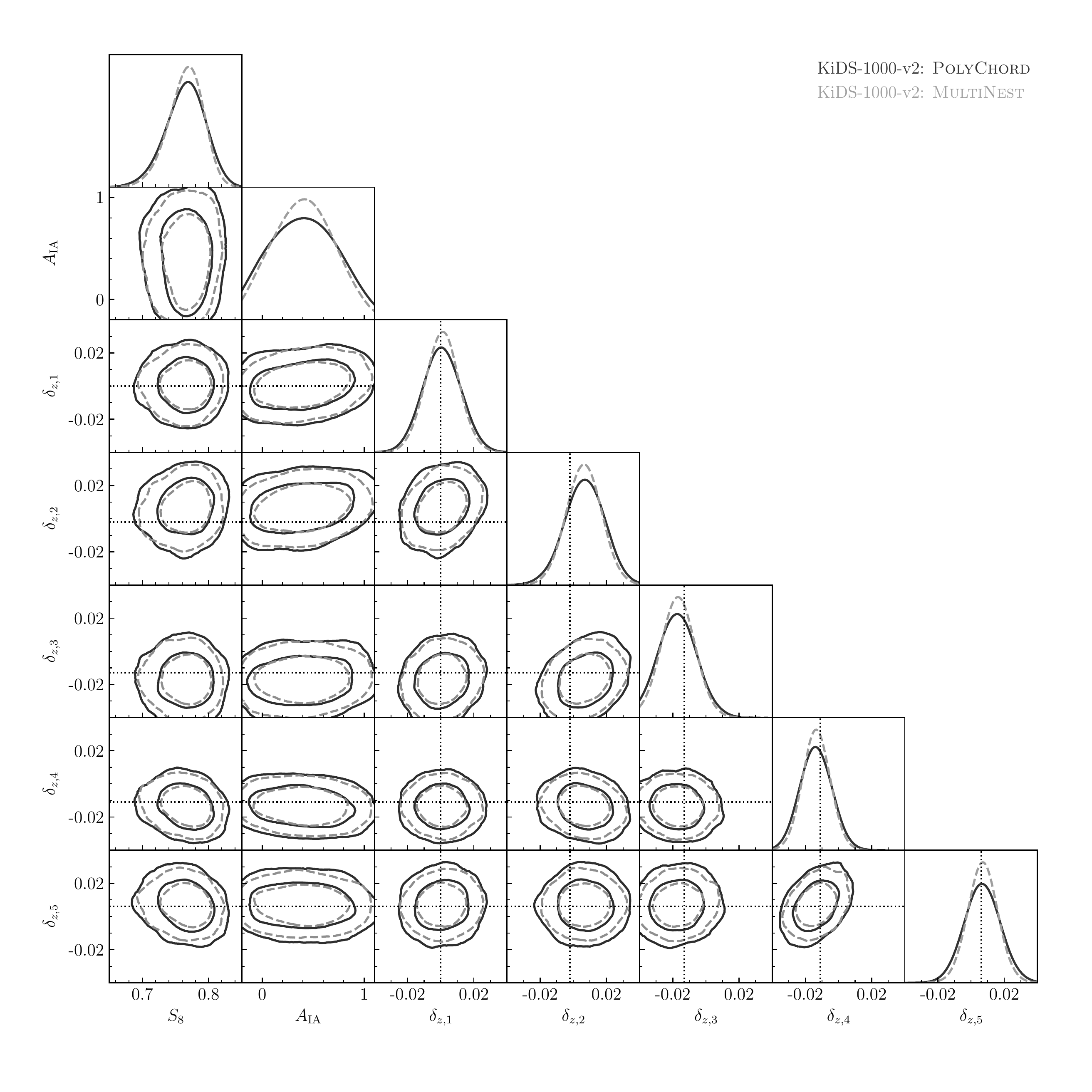}
   \caption{Posterior distributions of $S_8$ and nuisance parameters from our fiducial analyses, as generated by \textsc{PolyChord} (solid black lines) and \textsc{MultiNest} (dashed grey lines). The contours represent the $68\%$ and $95\%$ credible intervals and are smoothed using a Gaussian KDE with a bandwidth scaled by a factor of 1.5. We note that the Gaussian priors we have set, as outlined in Table~\ref{tab:priors}, strongly influence the redshift offset parameters. The dotted lines represent the central values of these Gaussian priors.}
    \label{fig:S8nuisance}
    \end{figure*}

\end{appendix}

\end{document}